\documentclass[]{article}

\usepackage[utf8]{inputenc}
\usepackage{graphicx}
\usepackage{authblk}
\usepackage{xr}
\usepackage{mwe}
\usepackage[dvipsnames]{xcolor}
\usepackage{hyperref}
\usepackage{ulem}
\usepackage{amssymb}
\usepackage{wasysym}
\usepackage{rotating}
\title{Surface Patterns Shaped by Additives in Crystals}
\author[]{M. A. Chabowska}
\author[]{M. A. Za\l uska-Kotur}

\affil[]{Institute of Physics, Polish Academy of Sciences, al. Lotników 32/46, Warsaw, Poland}

\date{\today}

\begin{document}
\maketitle


\begin{abstract}
One technique for creating semiconductor crystals with new, desired properties involves replacing some atoms in the crystal lattice with additives — atoms of a different type. This substitution not only alters the bulk properties of the crystal but also affects the patterns formed on its surface. A surface that is smooth and regular in a uniform crystal can become bunched or meandered under the same growth conditions if some atoms are replaced by additives. The Vicinal Cellular Automaton (VicCA) model is used to study this behavior, analyzing the mechanism of pattern formation when additives are introduced into the system. It has also been shown that the newly formed structures resulting from the presence of additives can be smoothed by  applying successive layers of a homogeneous composition  on top for a sufficiently long time. Additives can also act as smoothing agents for bunched or meandered surface patterns that develop in a homogeneous crystal. However, their effectiveness diminishes for spatially extended patterns such as nanowires. Typically, when homogeneous crystal layers are applied to spatial structures, the resulting surface is not entirely smoothed but instead transforms into a distinct shape. This phenomenon is demonstrated, and its underlying mechanism is thoroughly analyzed.
\end{abstract}

\section{Introduction}

The development of nanotechnology, the application of new techniques, and the design of novel nanodevices require a deeper understanding and control of the processes occurring during crystal growth, particularly in terms of surface dynamics. Recently, a significant driving force behind this research has been the search for a solid platform for quantum computing with topological protection \cite{oreg,lutchyn,mourik}. Extremely precise operations are necessary to prepare structures that will form the basis for building quantum computers, enabling the storage and processing of information recorded in qubits. Other examples of emerging nanoscale technologies include giant magnetoresistance (GMR) \cite{grunberg,baibich,fermon}, light-emitting diodes (LEDs) \cite{zeludev}, and electronic circuits based on memristors \cite{yang1,yang2}. The surface of crystals serves as the primary front for their growth, and its shape and stability during evolution determine both the speed and quality of crystal growth. This surface is where the final structure of the growing crystal, as well as the number of defects and incorporated impurities, are established.

One of the key goals in the electronics industry is to avoid rough surfaces in grown crystals, thereby achieving narrow interfaces between the layers of semiconductor structures. Wide interfaces degrade the quality of these structures and, consequently, the performance of semiconductor devices. Crystals used in electronic device production are expected to have regular structures, such as evenly spaced steps. As a result, determining the regions in the parameter space where smooth structures appear and identifying factors to avoid is an area of ongoing research. Conversely, the popularity of reconstructed but regular surface morphologies, such as step bunches or meanders, has recently increased due to their use as templates for the growth of nanowires — low-dimensional structures that are gaining widespread interest \cite{verre2012,yao2016}. The mechanisms of surface evolution, the causes of pattern formation, and the identification of growth conditions that ensure smooth or patterned morphologies are subjects of both experimental \cite{usov2011,shen2007,omi2005,ass_turski,CGD_sawicka} and theoretical \cite{liu1998,pimpinelli2002,krug2005,tonchev2010,stoyanov1997,fk2017_2,RMP_Misbah,Krukowski22} research. Bunched structures can emerge as a result of surface evolution in the presence of a driving force \cite{stoyanov1997} or be induced by an inverse Ehrlich-Schwoebel barrier (iES) \cite{fk2017_2,RMP_Misbah,Krukowski22,schwoebel1969, sato2001,xie2002}, which can be explained by the attachment of particles to a step originating from the upper rather than the lower terrace. It is easy to show that in such a situation, the regular pattern of equidistant steps becomes unstable, making step bunching highly likely. Another aspect of this instability is the meandering of steps, which forms on the surface of a growing crystal when a direct Ehrlich-Schwoebel barrier (ES) is present \cite{ass_turski,CGD_sawicka,fk2017_2,RMP_Misbah,schwoebel1966,JAP_krzyzew}. Both bunched and meandered patterns can serve as templates for nanostructure growth. Nanowires (NWs) appear to grow only on substrates with specific characteristic features. J. Kang et al. \cite{JK_NL} demonstrated experimentally and theoretically that gold droplets start to nucleate and guide NW growth only when the \{111\}B facets become sufficiently large and regular. Results from Monte Carlo simulations show that a minimal stepless region is required to maintain supersaturation conditions in the Au droplet that initiates NW growth. The authors concluded that the substrate surface morphology, including the \{111\} facet structure, plays a key role in the NW crystallization process. Understanding surface dynamics also enables better control over the growth of structures with desired geometry and properties. For example, as shown in Ref.~\cite{Arora}, highly regular, periodic step-bunched surfaces of n-type doped vicinal Si(111) were used to obtain regular planar Co NW systems.

From our perspective, the most intriguing aspect is the analysis of vicinal surfaces. Previous studies have primarily focused on various aspects of crystal growth \cite{AIP_Krasteva,Sudoh_2003,Neel_2003,Rahman_2003,Minoda_2003,Rousset_2003}, including step bunching and meandering \cite{JCG_krzyzew,prb_Toktarbaiuly,CGD_krzyzew,CGD_popova}. Specifically, the emergence of step bunching instability has been systematically investigated \cite{JCG_krzyzew,CGD_popova,Popova-CGD23}. Using the Vicinal Cellular Automaton model, F. Krzyżewski et al. \cite{JCG_krzyzew} examined the stability of step bunching during one-dimensional (1D) growth and sublimation of vicinal surfaces under two destabilization modes — step-down and step-up currents. They demonstrated that it is possible to reproduce step bunching instability caused by two opposite drift directions in a system where sublimation balances growth. The step bunching process has also been reported for two-dimensional (2D) surfaces.  The step  bunching  process has also been reported for 2D surfaces \cite{crystals_MZK}. It has been shown that simply controlling the presence and height of direct and inverse Ehrlich–Schwoebel barriers and properly selecting the well potential between them leads to the growth of nanocolumns, nanowires, and nanopyramids or meanders in the same system. Another study focused on the two-dimensional vicinal surface, revealing that the formation of different structures is not only influenced by the combination of step barriers but also by the diffusion rate \cite{Chabowska-ACS}. We presented an analysis of how diffusion rate affects surface patterns, including changes in surface structure features. Step bunches, meanders, nanowires, islands, and mounds of various shapes are structures sought after as foundations for implementing new, technologically significant concepts. To achieve the desired surface shape, different types of internal properties at the nanoscale must be considered.

In this study, we analyze how the dependence of pattern formation on crystal composition or surfactant presence can be modeled through the proper choice of the potential energy  landscape. There are numerous examples of significant changes in surface structure — sometimes dramatic — as a consequence of modification in crystal composition, even when the crystal structure itself remains unchanged.
As demonstrated by A. Khokhryakov et al. (Ref. \cite{Exp1}), even a minimal amount of silicon present within the Mg-C structure can result in the formation of macrosteps. Furthermore, as evidenced in Ref. \cite{Exp2}, an elevated concentration of silicon impurities within $ \beta-\rm{Ga_2O_3}$ can transform the step-flow morphology into the characteristic features of step-bunching morphology. By adjusting the growth conditions, the authors were able to recover the step-flow morphology. The morphological evolution of the GaSb crystal in the absence and presence of Bi surfactant was investigated in Ref. \cite{Exp3}. The results presented there demonstrate that the addition of Bi can prevent the formation of undesired 3D islands, lead to anti-step-bunching effects, and promote the transition from step-meandering to mound morphologies. These findings, along with those reported in Ref. \cite{Exp4} for AlN/AlGaN superlattices, highlight the ability of surfactants to improve surface morphology. Moreover, additives have been shown to influence the stability and growth of various crystallographic planes of growing crystals, as demonstrated in Ref.\cite{Exp5} for $\rm{Cu_2O}$ crystals. If additives are broadly interpreted, the ternary system examined in Ref. \cite{ternary} could also serve as an example of the unexpected morphological variability of InP(As) nanostructures.

The theoretical analysis of such phenomena usually relies on variations in interaction strengths between different types of atoms, differences in diffusion rates, or the immobility of impurities. While these approaches have successfully explained many aspects of the experimental findings (Ref. \cite{Th1,Th2,Th3,Th4,Th5}), they often required a distinct model for each case.
In this work, we analyze such behavior from the perspective of the potential energy landscape experienced by diffusing particles. We demonstrate that local modifications to the potential can reproduce a variety of surface patterns, closely resembling those observed experimentally. Furthermore, we show that significantly different system evolutions can emerge within the same unified framework.

We assume that an atom of a different type, when placed in a lattice site, introduces an additional potential into the system, thereby influencing the diffusion of adatoms along the surface for all atoms and leading to the formation of specific patterns. We discuss three examples of such behavior. In the first example, additives in the system induce iES barrier near their location, causing step bunching. In the second example, ES barrier appear due to the presence of additives on the steps, leading to step meandering. Interestingly, both of these patterning processes can be reversed by growing layers of homogeneous composition on top of the pattern for a sufficiently long time. In cases where three-dimensional (3D) structures such as nanowires or mounds form on the surface, smoothing the surface is more challenging, though still possible to some extent, as illustrated in the third example. We demonstrate one possible mechanism by which additives influence pattern formation. Several other potential consequences of adding different particles to the system can be analyzed in a similar manner. The approach presented here for analyzing pattern formation in binary or ternary crystals is quite general and can be applied as is or adapted for other scenarios.

\section{Model}
The model which we use in this work is (2 + 1)D vicinal Cellular Automaton model, introduced and studied before in various (1 + 1)D contexts
\cite{AIP_Krasteva,JCG_krzyzew,prb_Toktarbaiuly,CGD_krzyzew,CGD_popova,Popova-CGD23} and (2 + 1)D context \cite{crystals_MZK,Chabowska-ACS, Chabowska-PRB}. It is built as a combination of two essentially different modules: the Cellular Automaton (CA) one responsible for the evolution of the vicinal crystal surface and the Monte Carlo (MC) one representing the diffusion of the adatoms. The CA module realizes the growth of the surface according to pre-defined rules in a parallel fashion while MC module realizes the diffusion of the adatoms in the serial mode, adatom after adatom.

One diffusional step is completed when each adatom is visited once (on average). A single time step of the simulation is represented by the diffusion of all adatoms along the surface (MC unit), then one growth update (CA unit) and finally compensating the adatoms to their initial concentration $c_0$. This design allows the study of large systems in long simulations. Between two growth modules, all adatoms try $n_{DS}$ diffusional jumps, but only those that point to neighboring unoccupied lattice site are made. As $n_{DS}$ grows, it goes from diffusion-limited growth (DL) towards kinetic-limited growth mode (KL) and at the same time the transparency of the step increases.

The system  is  described  by  two  elements — the surface of crystal represented by a table with the height of the crystal, given by the number of build-in atom layers, and the second part — the layer of adatoms. The surface usually consists of descending steps. They fall from left to right and are initially separated by $l_0$ length terraces. In the direction along the steps, periodic boundary conditions are imposed, while across steps, helical periodic boundary conditions are applied to maintain the step differences.

The CA rules determine when an adatom is incorporated into the crystal. There are three different situations where an adatom becomes part of the crystal:
one when adatom is  at  kink  position —  it  is  adsorbed  unconditionally. Second is  for  atoms  that are at straight steps.  They  are adsorbed  into  the  crystals,  but  with  a lower  probability.
Such  lower  probability  in  the realization  used  below  is provided  by  the rule  that at  least one  neighboring site is  also occupied. We assumed that the particles are easily built in the crystal at kinks, and more difficult at the straight part of the step. The step stiffness can be regulated, making a second event to be more or less likely. A step is stiffer when it is harder to build an adatom into the straight step. And  the  third  possibility — island
creation, it is adsorption of the atom if three or more adatoms are neighboring. This last conditions is sometimes weakened by demand of four neighboring atoms.
More details about the model are described in Ref. \cite{crystals_MZK,Chabowska-ACS, Chabowska-PRB}.

\begin{figure}[hbt]
 \centering
\includegraphics[width=0.32\textwidth]{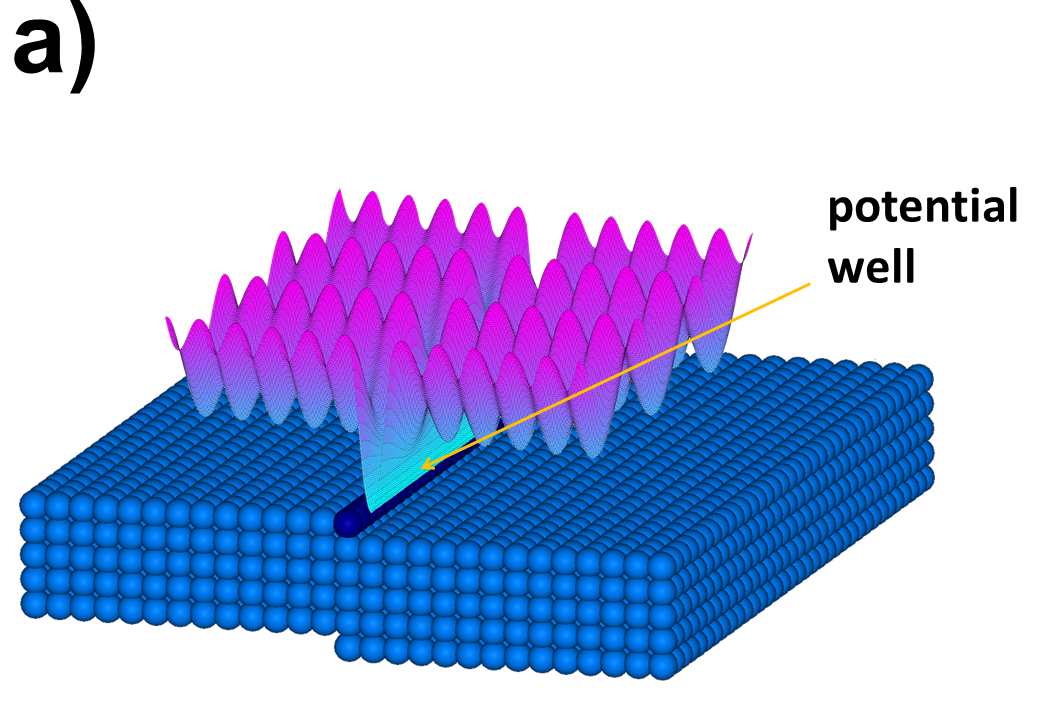}
\includegraphics[width=0.32\textwidth]{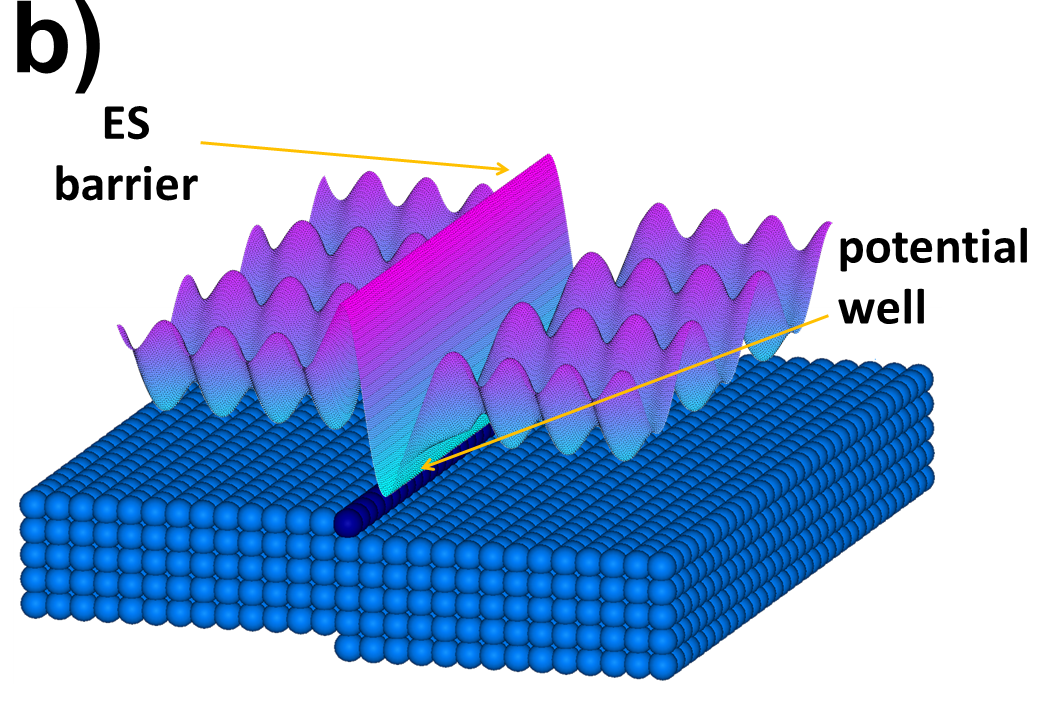}
\includegraphics[width=0.32\textwidth]{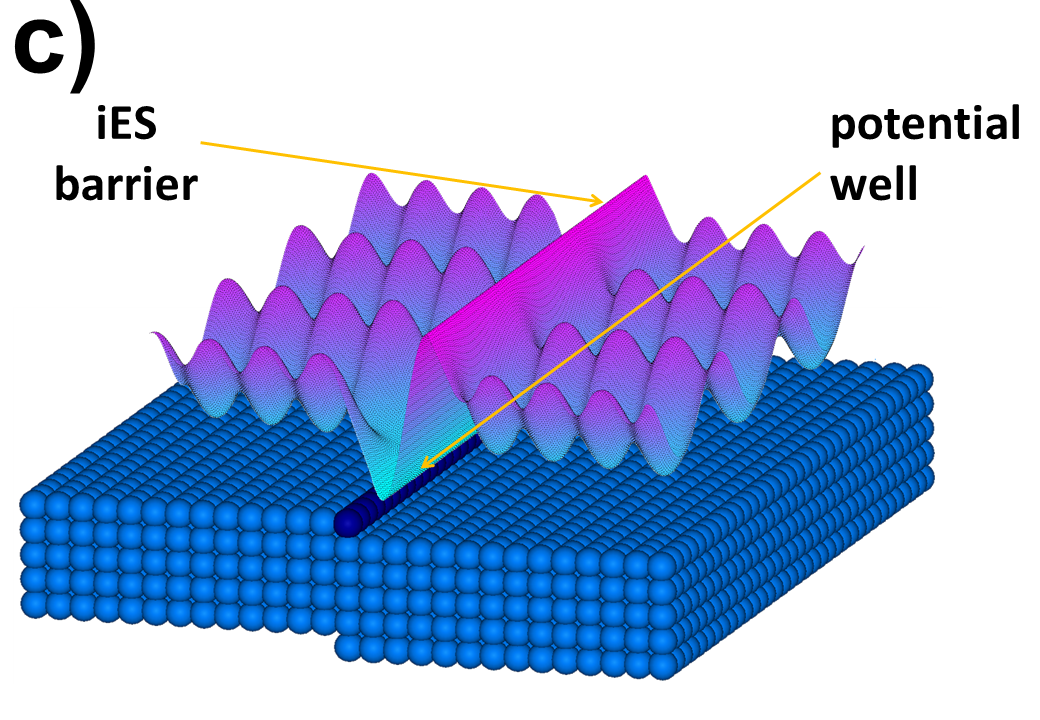}
\caption{Surface with descending step from the left to the right and visualization of the energy  potential landscape on top of the surface (a) potential well at the bottom of step, (b) potential well and Ehrlich–Schwoebel, (c) potential well and inverse Ehrlich–Schwoebel barrier.}
\label{fig:init}
\end{figure}

A very  important  element  of  the   model  is  energy  potential  landscape in which all atoms diffuse. Its structure, in particular some important elements of it decide about the formed surface pattern. Three examples of  such potentials are  shown in Figure~\ref{fig:init}. We  assume that all jumps along the terraces, except for those in the immediate vicinity of the step, are performed with the same probability, which is equal to 1 after the equal choice of jump direction. Sites at the bottom of step have different energy, this is energy  well related to the interaction energy with the step \cite{akiyama1,akiyama2,akiyama3}. Such situation is illustrated in Fig.~\ref{fig:init}a. The surface energy landscape in general can have very complex shape, it depends on many different factors, like lattice deformations, surface reconstruction, and all broken bonds at each crystal surface. Due to this there are in this landscape not only potential wells, but also barriers of different types, among them two are the most important ones, having the largest consequences. That are the direct ES barrier at top of the step  —  illustrated in the Fig.~\ref{fig:init}b and the inverse iES barrier at the bottom of steps.
They make jumping across the step in the first case or towards step in the  second more difficult. We set the probability of such jump $P_{ES}=\exp(-\beta E_{ES})$ and  $P_{iES}=\exp(-\beta E_{iES})$,  where $\beta=1/(k_BT)$ is  temperature  factor  and  $E_{ES}, E_{iES}$ are energies of barriers that have to be overpass by jumping atom. Similarly, probability of jumping out from the  potential well is equal to $P_V=\exp(-\beta E_{V})$ with $E_V$ being the energy difference between of potential well and average energy level.

\begin{figure}[hbt]
 \centering
\includegraphics[width=0.52\textwidth]{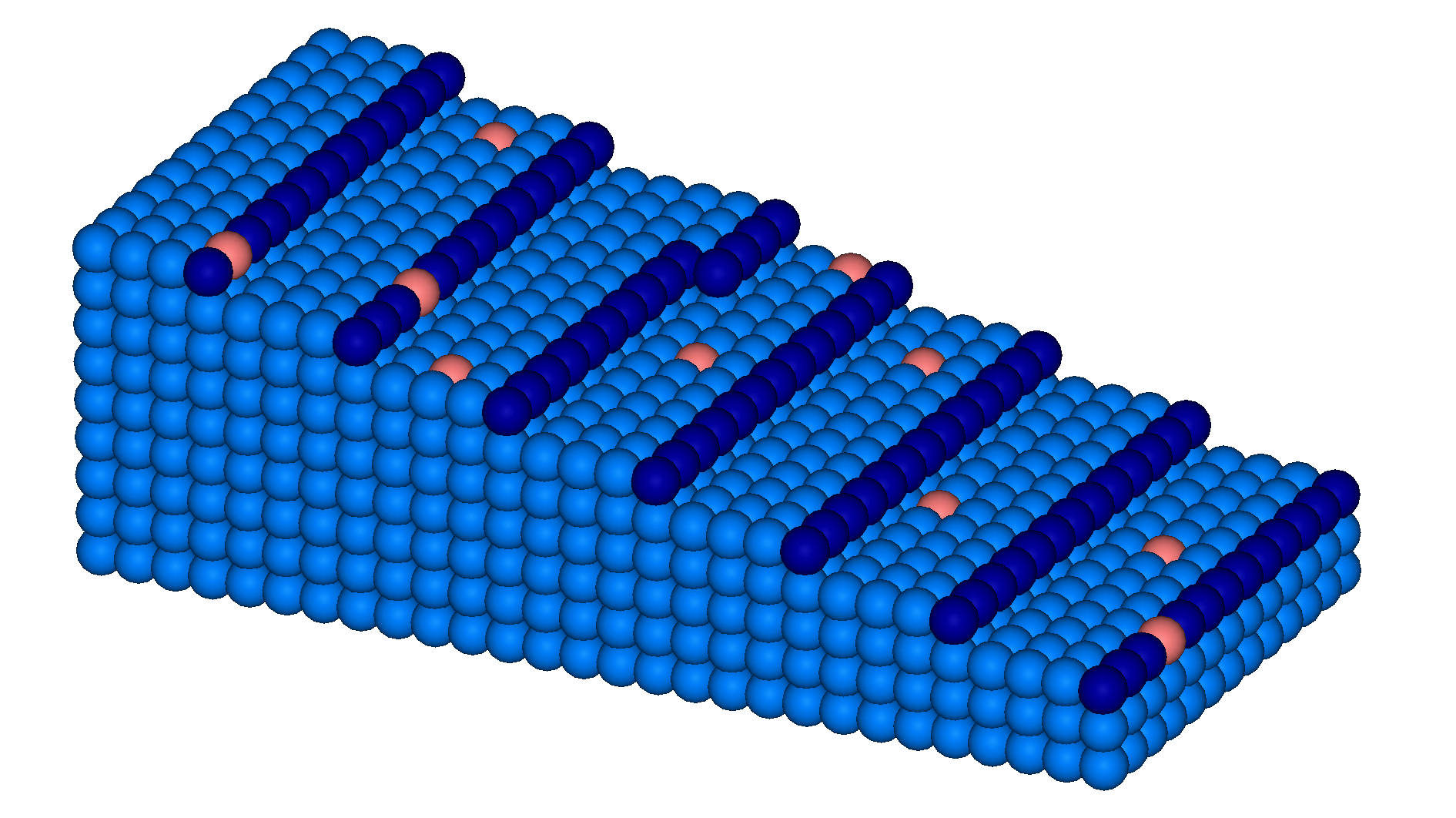}
\caption{Surface with descending steps from the left to the right with marked in red impurities or additives.}
\label{fig:additives}
\end{figure}

Usually, iES barrier together with potential well are responsible for the step bunching, while the  presence of ES barrier causes meandering process. The aim of our study is to check how additives introduced to the system in binary or ternary crystal can be  responsible for step bunching or meandering.
Fig.~\ref{fig:additives}.

To model such situation we prepared system with such surface potential that crystal with it grows smoothly and regularly. It means regular smooth potential with small energy well of order $\beta E_{V1}=1$. Such factor means that at temperature of about 700~K the energy is 0.064~eV.
\begin{figure}[hbt]
 \centering
\includegraphics[width=0.32\textwidth]{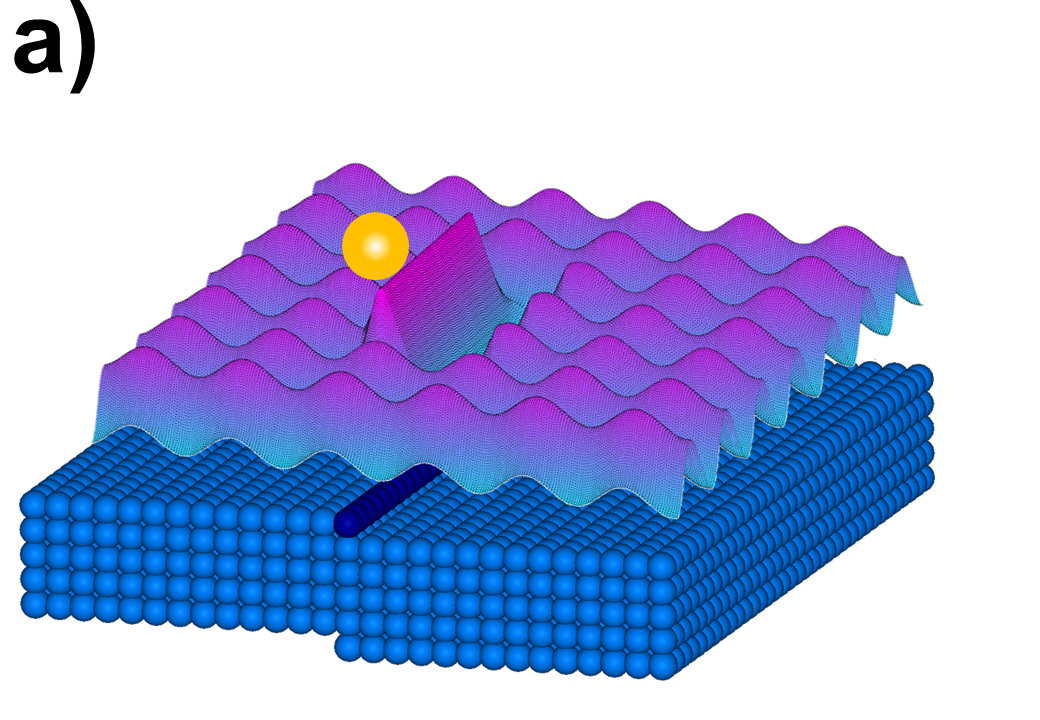}
\includegraphics[width=0.32\textwidth]{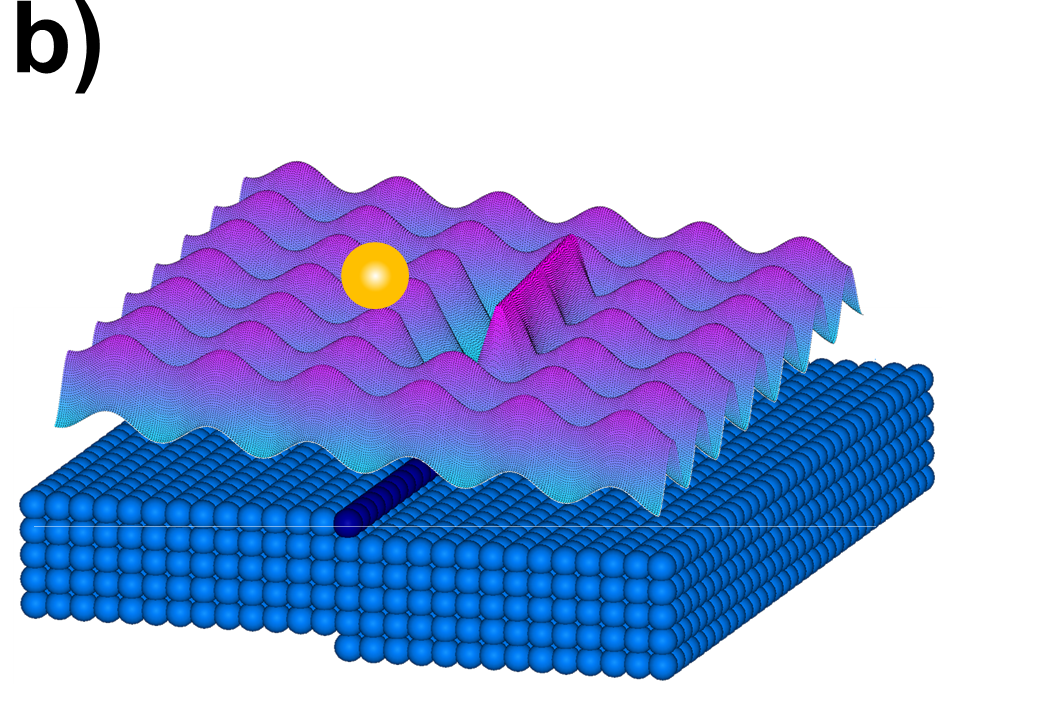}
\includegraphics[width=0.32\textwidth]{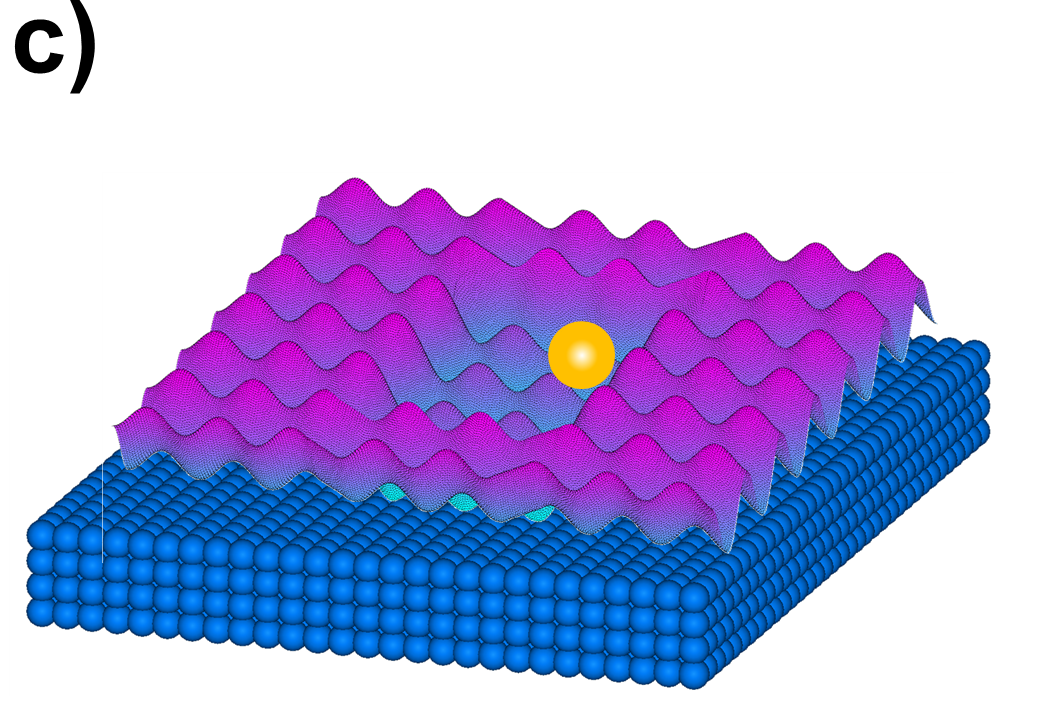}
\caption{Local modification of energy surface potential caused by the presence of additive at the step. (a) Potential well with ES barrier, (b) potential well with iES barrier and (c) potential well on top of nanowire.}
\label{fig:init_additives}
\end{figure}
Then  additives of given concentration replaced atoms at randomly chosen lattice sites (see Fig.~\ref{fig:additives}). We assume that each of such atoms modifies energy landscape in the way shown in Fig.~\ref{fig:init_additives}. It means that when the additive is placed at the step, an additional potential well appears below it, extending to $k$ neighboring sites in both directions along the step. This additional potential well is accompanied by ES barrier in the first case (Fig.~\ref{fig:init_additives}a) and iES barrier in the second case (Fig.~\ref{fig:init_additives}b). We see that both, energy well and energy barrier spread out by three consecutive sites ($k=1$).

The next case illustrated in Fig.~\ref{fig:init_additives}c shows the energy well, which causes the molecules to gather inside and increases the probability of nucleation. Then, if this potential appears at the top of the cluster, it becomes the seed for the growth of a nanowire. We will demonstrate such 3D growth in the third example below.

\section{Results}
\subsection{Step bunching and meandering}
The presence of an inverse barrier at the step is one of the possible and most obvious reasons for the formation of step bunches in the system \cite{krug2005,fk2017_2,RMP_Misbah,sato2001,JCG_krzyzew,CGD_krzyzew,Chabowska-ACS}. It is an energy barrier, occurring at the bottom of the step, one lattice site away from it, as shown in Fig.~\ref{fig:init_additives}a. Such a barrier stops the flow of particles towards the step from the bottom, which, as can be easily shown by stability analysis, causes the step bunching. Using the VicCA approach, the influence of the inverse step barrier on the step bunching process has been investigated and discussed earlier under different conditions given by the model parameters in 1D \cite{CGD_krzyzew} and 2D \cite{crystals_MZK} systems. In this work, we show that step bunching is also induced  when iES barrier occurs only at the locations where the additives are attached and in their vicinity, and not along the entire step length.

\begin{figure}[hbt]
 \centering
a)\includegraphics[width=0.3\textwidth]{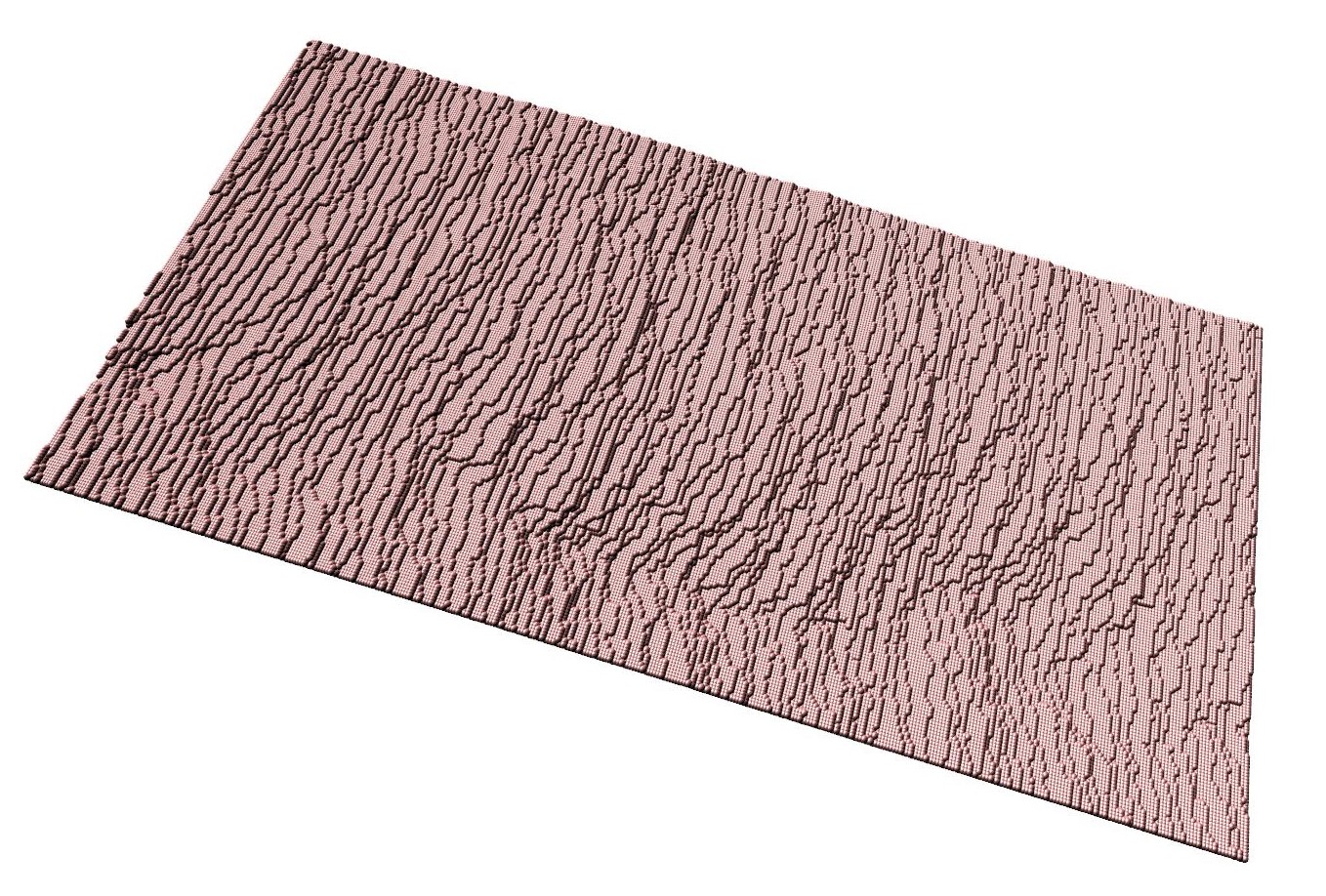}
b)\includegraphics[width=0.3\textwidth]{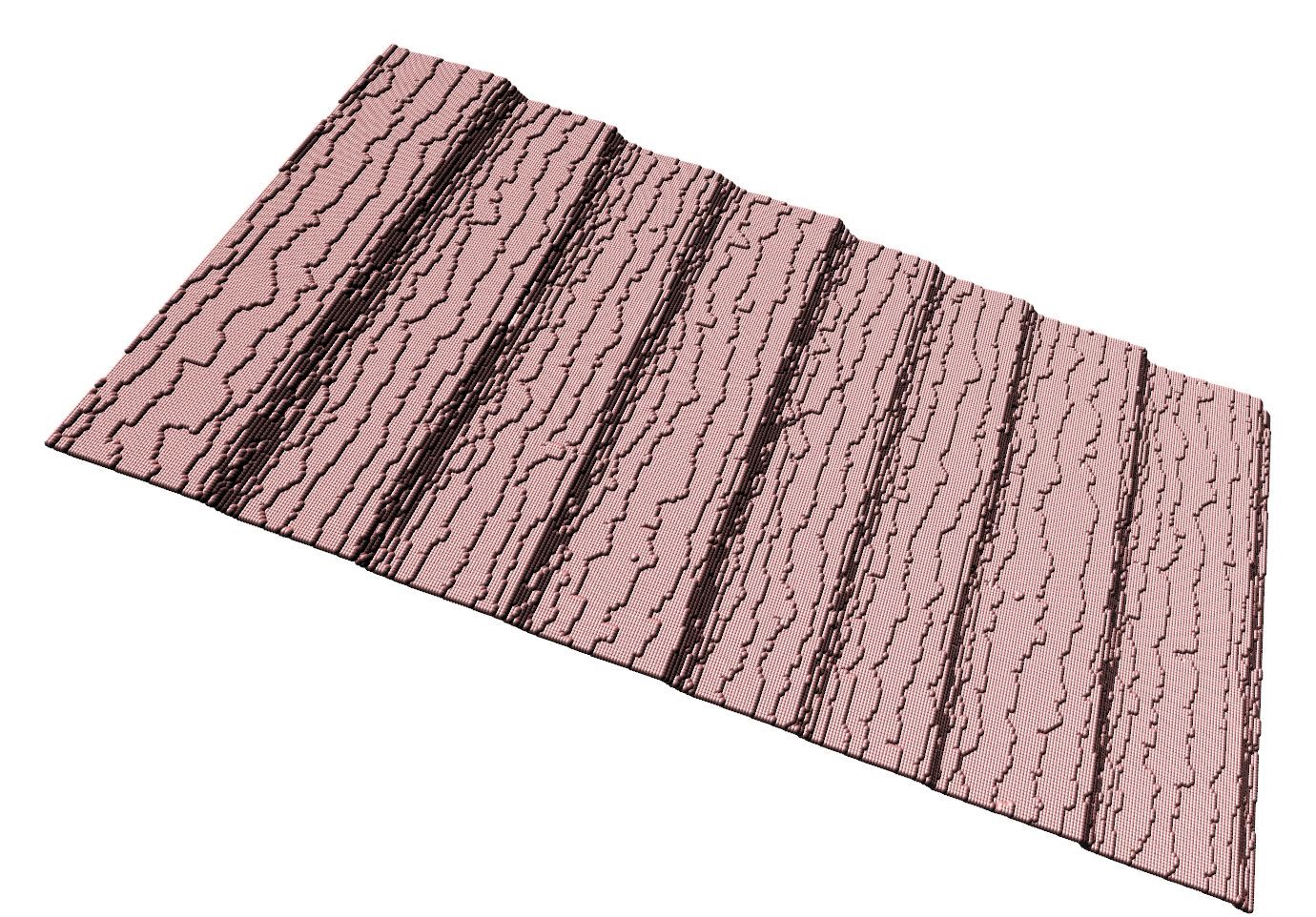}
c)\includegraphics[width=0.3\textwidth]{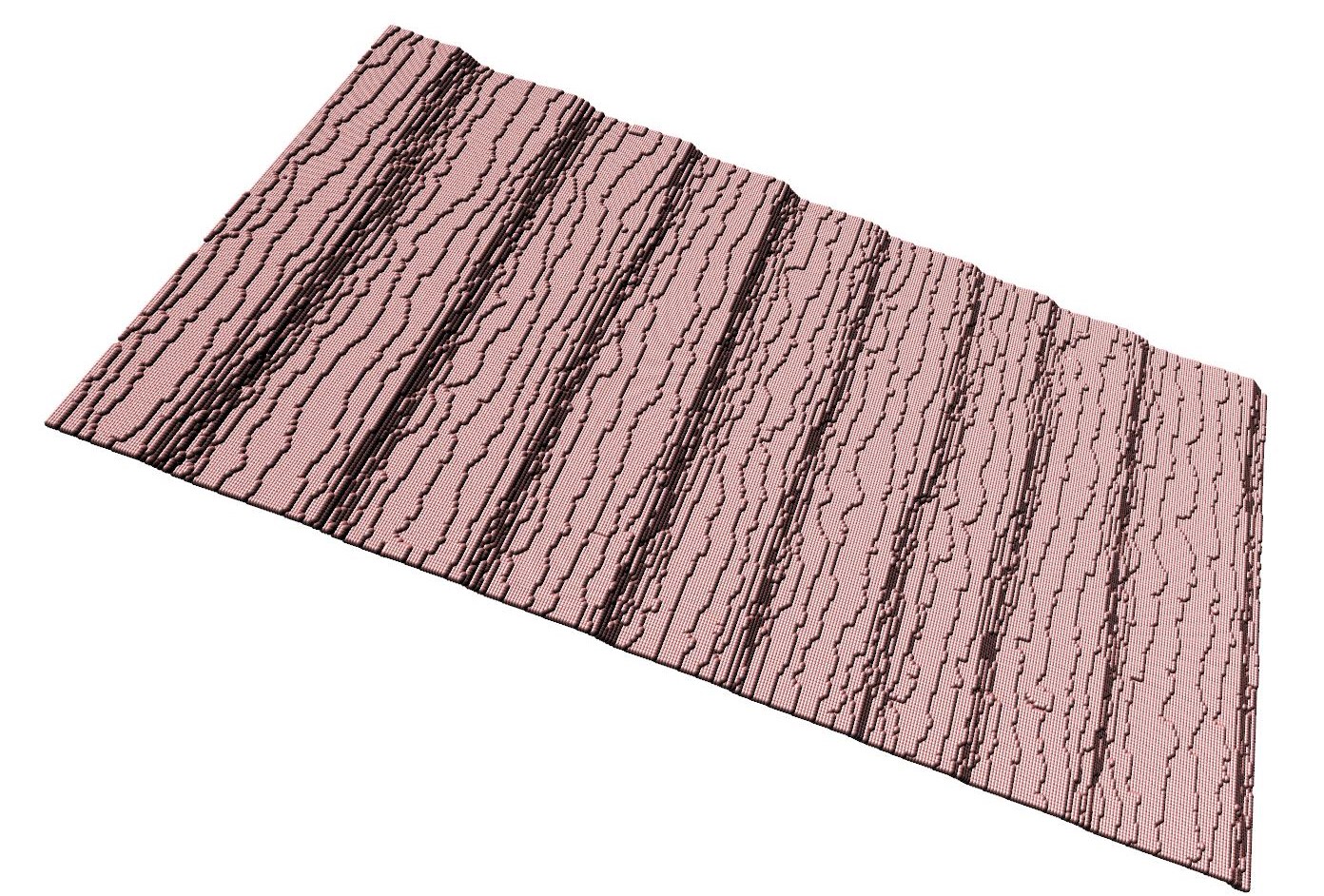}
\caption{Surface of crystal after $2 \cdot 10^6$ simulation steps. Each atom contributes a step potential $\beta E_{V1}=1$ with no additional barrier. Each additive atom contributes a potential $\beta E_{V2}=1.3$ and an iES barrier $\beta E_{iES}=6$, spreading along the step for $k$ sites on both sides of the additive. a) 0\% of additives, b) 30\% of additives with $k=2$, c) 3\% of additives with $k=15$.}
\label{fig:bunch_form}
\end{figure}
\begin{figure}
 \centering
a)\includegraphics[width=0.42\textwidth]{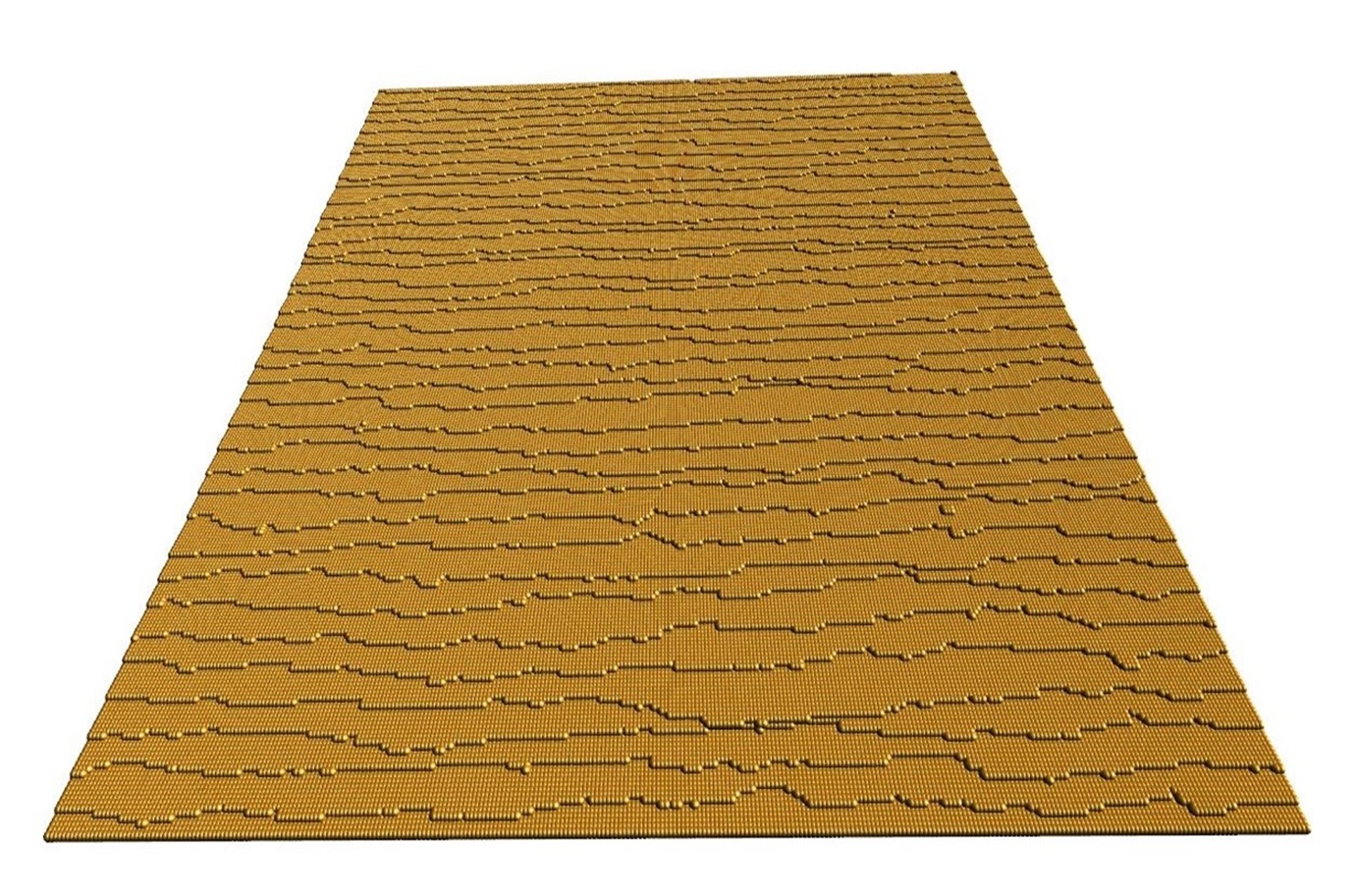}
b)\includegraphics[width=0.42\textwidth]{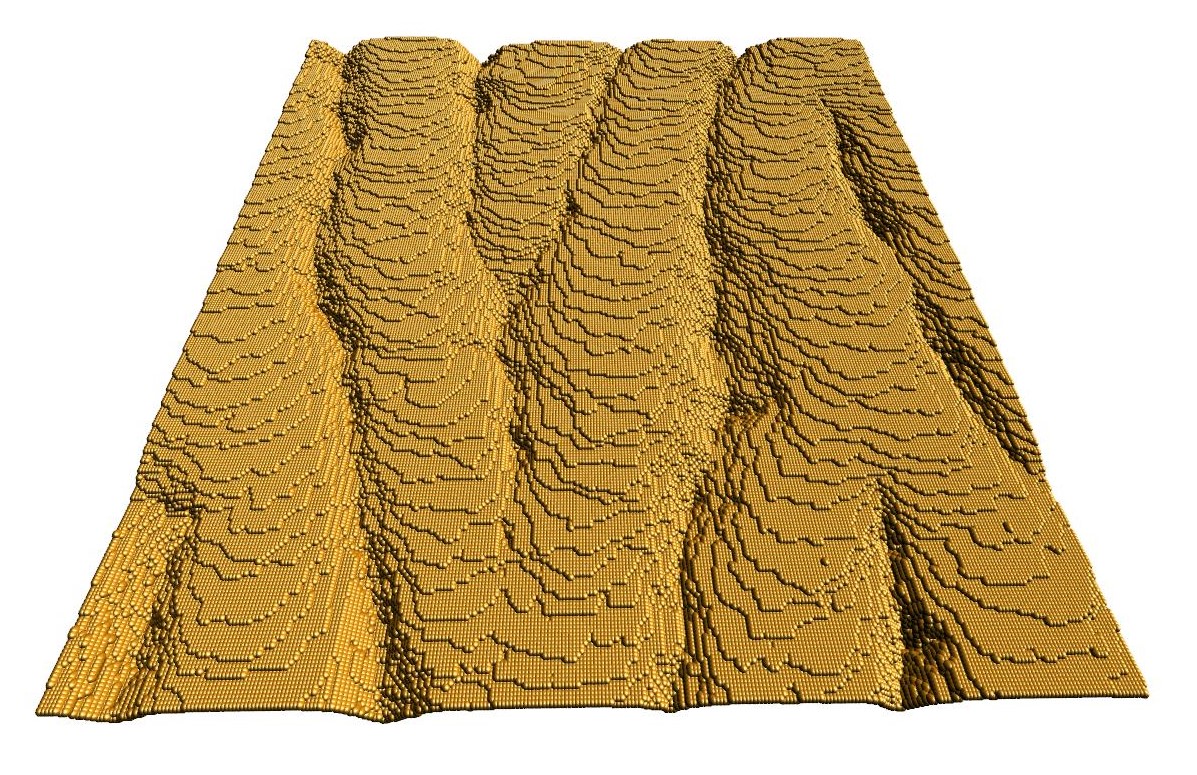}
c)\includegraphics[width=0.45\textwidth]{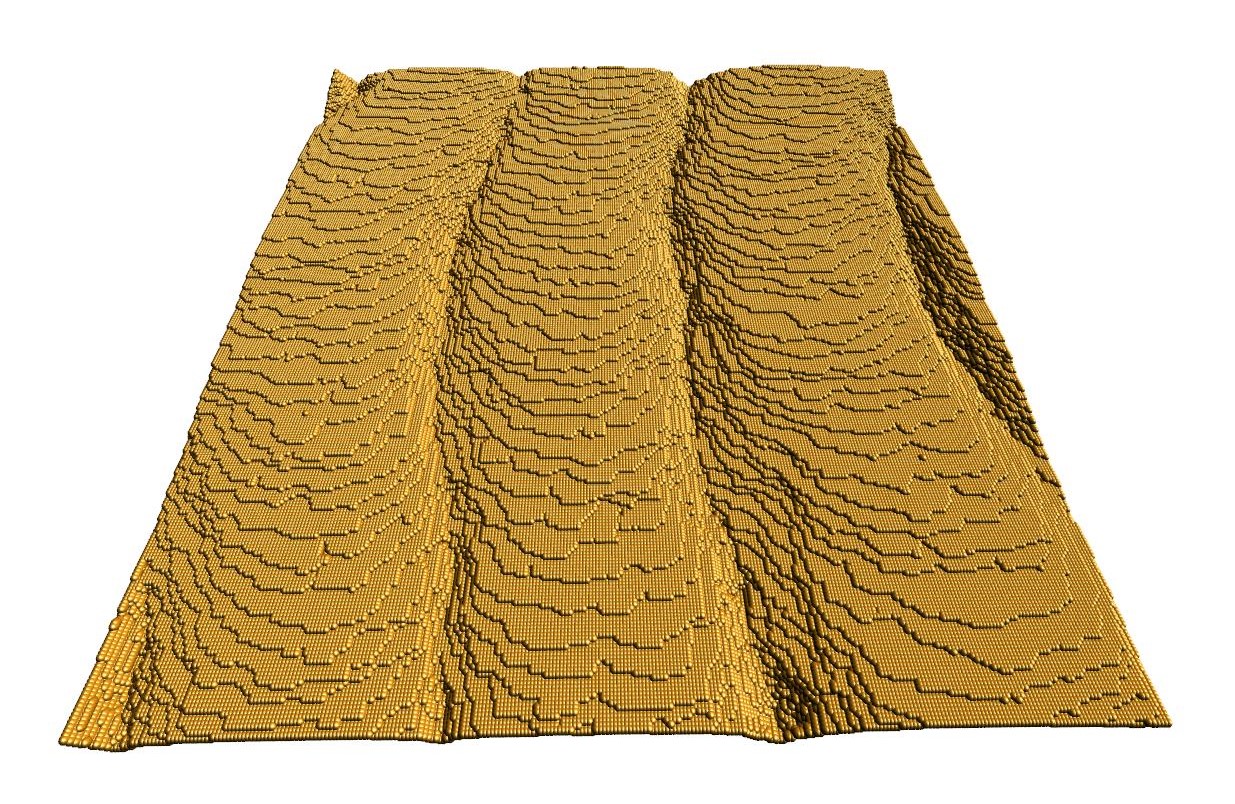}
d)\includegraphics[width=0.45\textwidth]{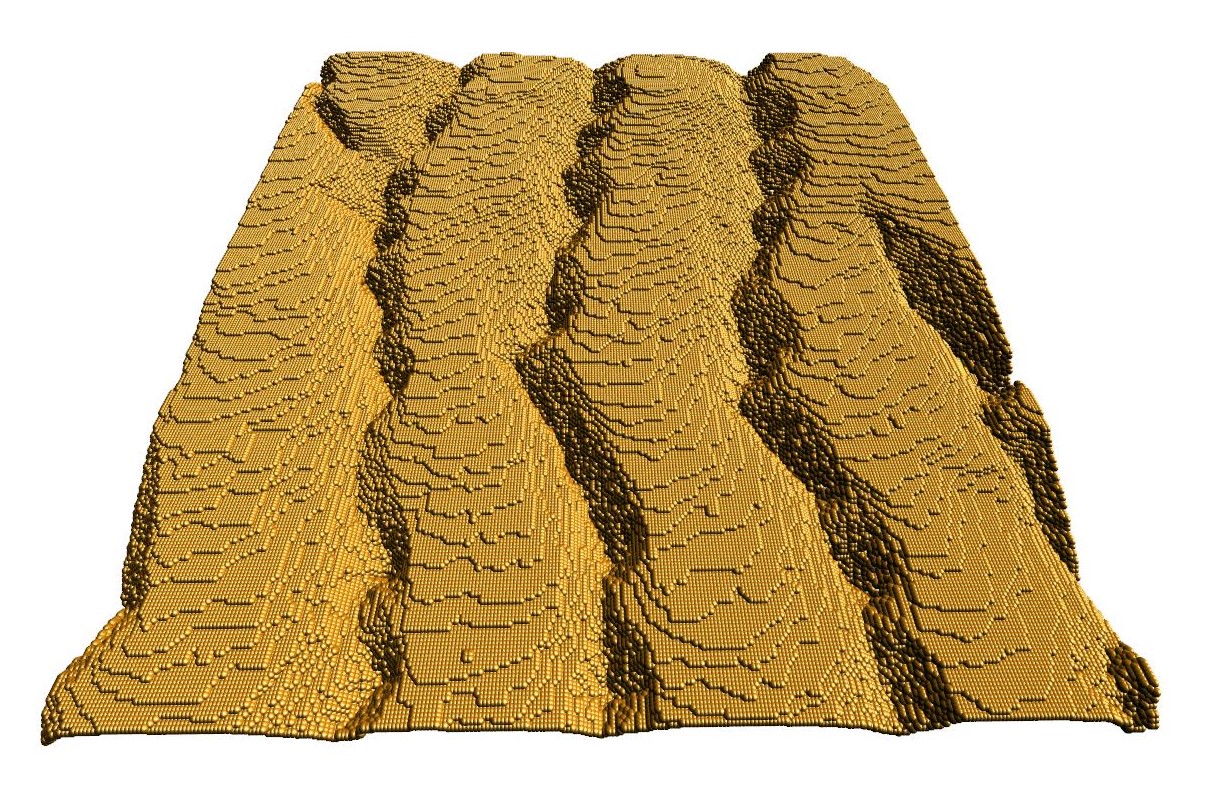}
\caption{Surface of crystal after $2 \cdot 10^6$ simulation steps. Each atom contributes a step potential $\beta E_{V1}=1$ with no additional barrier. Each additive atom contributes a potential $\beta E_{V2}=5$ and an ES barrier $\beta E_{ES}=6$, spreading along the step for $k$ sites on both sides of the additive. a) 0\% of additives, b) 20\% of additives with $k=1$, c) 5\% of additives with $k=3$, and d) 1\% of additives with $k=15$.}
\label{fig:meander_form}
\end{figure}
\begin{figure}
 \centering
a)\includegraphics*[width=0.42\textwidth, angle=90]{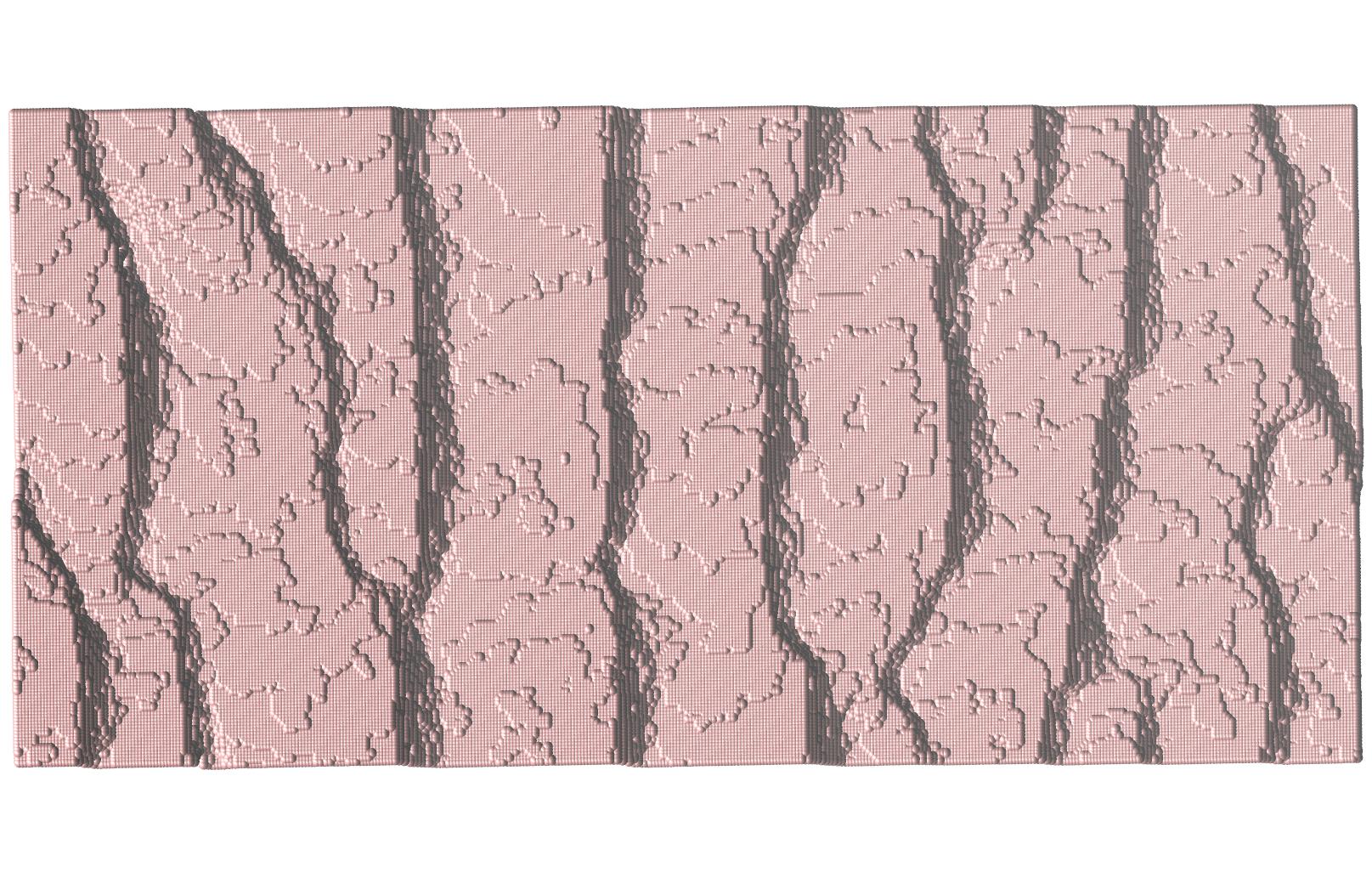}
b)\includegraphics*[width=0.42\textwidth, angle=90]{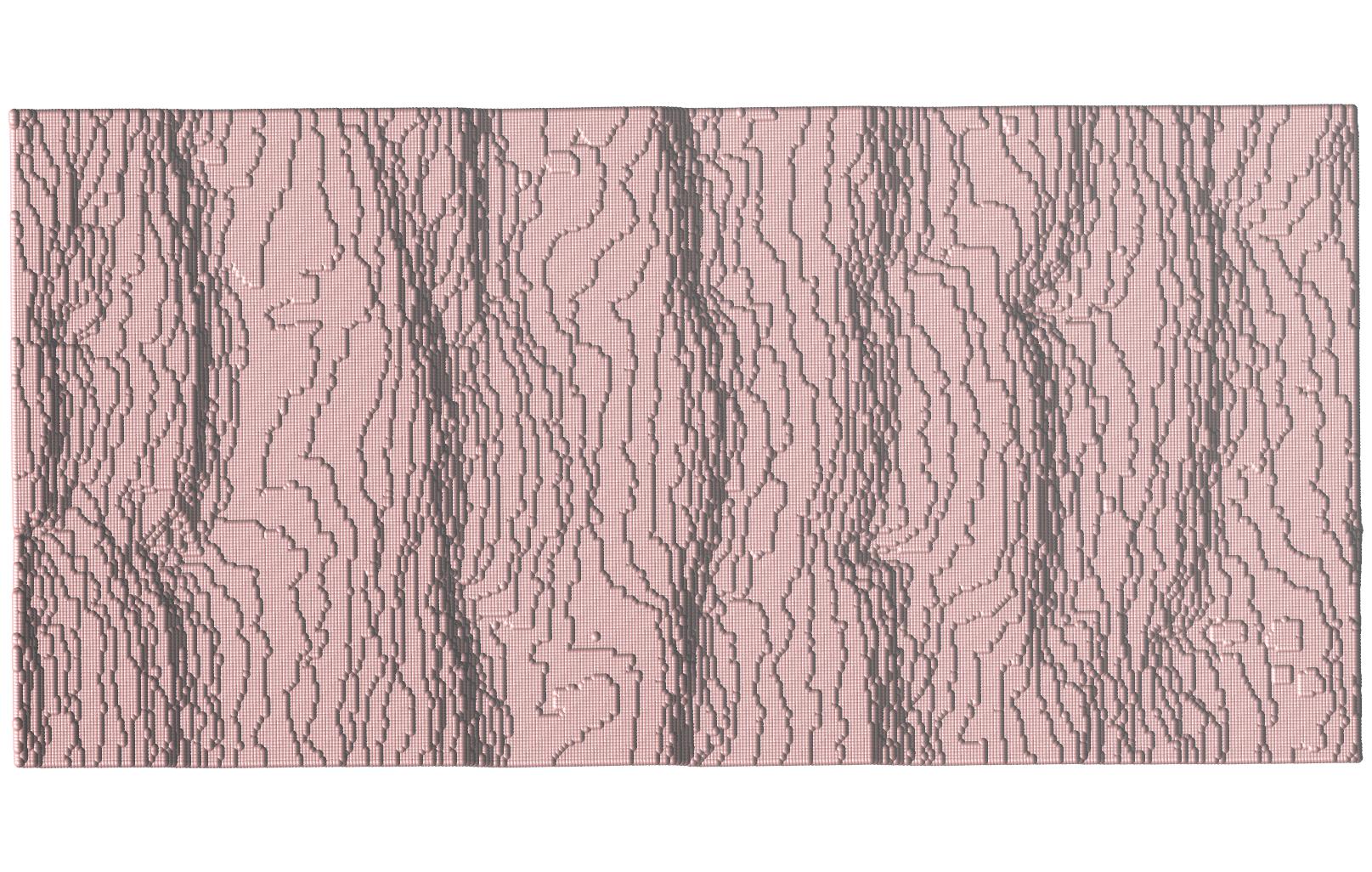}
c)\includegraphics*[width=0.42\textwidth, angle=90]{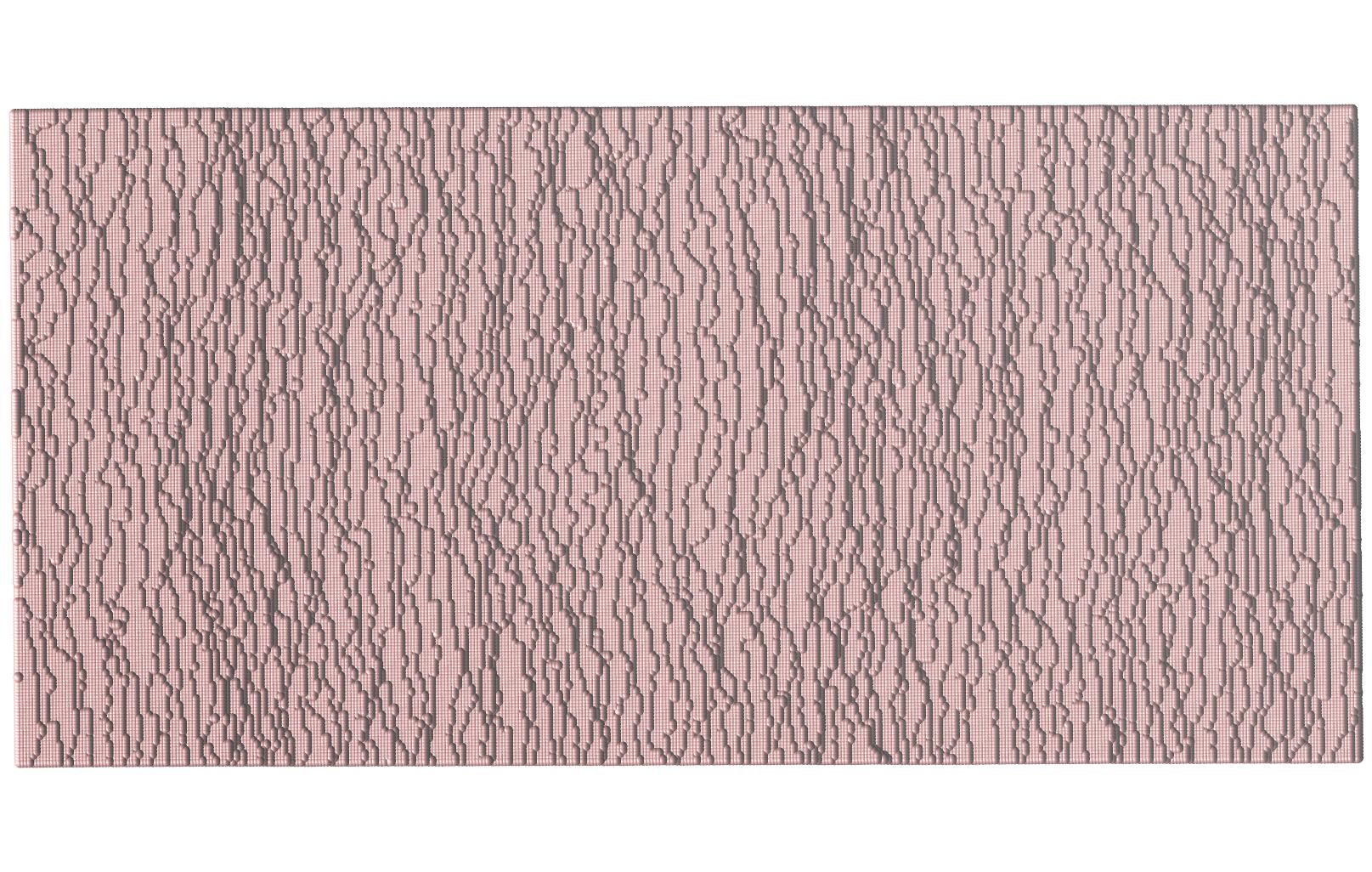}
\caption{Debunching process: a) The system after $10^6$ growth steps with two types of atoms in the incoming flux, where 30\% of the stream consists of  additives that introduce iES barrier into the system, b) after $10^5$ evolution steps with only neutral atoms delivered in the flux; and c) after another $2 \cdot 10^6$ evolution steps with neutral atoms.}
\label{fig:bunch}
\end{figure}
The simulation typically begins with a regular array with uniformly spaced steps. In our system, we assume low barrier at the bottom of steps, $\beta E_{V1}=1$, no barriers of any type present. We started with this low potential value to ensure it would not cause any meandering and that the system would evolve smoothly without any additives, as shown in Figs.~\ref{fig:bunch_form}a and~\ref{fig:meander_form}a.
Next we randomly replaced 30\% lattice sites with other atoms. Each of these atoms, when positioned on a step, introduces small additional potential well with a depth $\beta E_{V2}=1.3$ and an inverse Schwoebel barrier of $\beta E_{iES}=6$. Both extend to the next nearest neighboring sites $k=2$, as shown in Figure~\ref{fig:init_additives}. Other parameters used in this simulation were $ l_0=4, c_0 = 0.02, n_{DS}= 3$, and a system size of $200\times400$. After $2 \cdot 10^6$ simulation steps, well-defined bunches were formed, as shown in Fig.~\ref{fig:bunch_form}b. It is clear that the presence of additives, each introducing potential well and an inverse Schwoebel barrier, causes step bunching in an otherwise unbunched system. To check if we can achieve the same result in a more diluted system, we used 3\% of additives. We found that, in this case, inducing bunching in the system required extending the potentials attributed to each additive over long distances. Otherwise, bunches do not form.
In Fig.~\ref{fig:bunch_form}c, we present the results obtained for $k=15$. While the patterns in Fig.~\ref{fig:bunch_form}b and Fig.~\ref{fig:bunch_form}c appear very similar, they were achieved through two distinctly different methods. In the second example, the low fraction of additives was compensated by the extended range of the potential associated with each additive. The next case we wanted to study is the opposite situation: the presence ES barrier not iES barrier in the system. This barrier is known to induce step  meandering at the surface \cite{ass_turski,fk2017_2,RMP_Misbah,JAP_krzyzew}. Again, we start with a system characterized by a low potential well at the step, $\beta E_{V1}=1$, and replace 20\%, 5\%, and 1\% of the lattice sites with additives. Each additive is associated with a potential well extending over three sites ($k=1$), seven sites ($k=3$), and thirty-one sites ($k=15$), respectively, with $\beta E_{V2}=5$ and an ES barrier of $\beta E_{ES}=6$. We observed that meanders are more readily induced by additives compared to bunches. Consequently, we employed smaller fractions of additives and limited the number of sites over which the potentials are extended. Other parameters used were $ l_0=8, c_0 = 0.01, n_{DS}= 3$, and the system size was $300\times 304$ with a simulation time of $10^6$ time steps. After this duration, we observe a well-formed meandered structure, which develops significantly faster in contrast to the bunching process induced by the iES barrier. In Fig.~\ref{fig:meander_form}a, we see the smooth surface of a system without any additives. This can be compared to three other cases where decreasing fractions of additives are accompanied by an increasing range of potential extension. These combined factors demonstrate their influence, resulting in well-defined, meandered patterns. Notably, with the chosen parameters, we achieved relatively similar meander wavelengths. However, other parameter combinations can lead to different outcomes.

\begin{figure}[hbt]
 \centering
a)\includegraphics[width=0.3\textwidth]{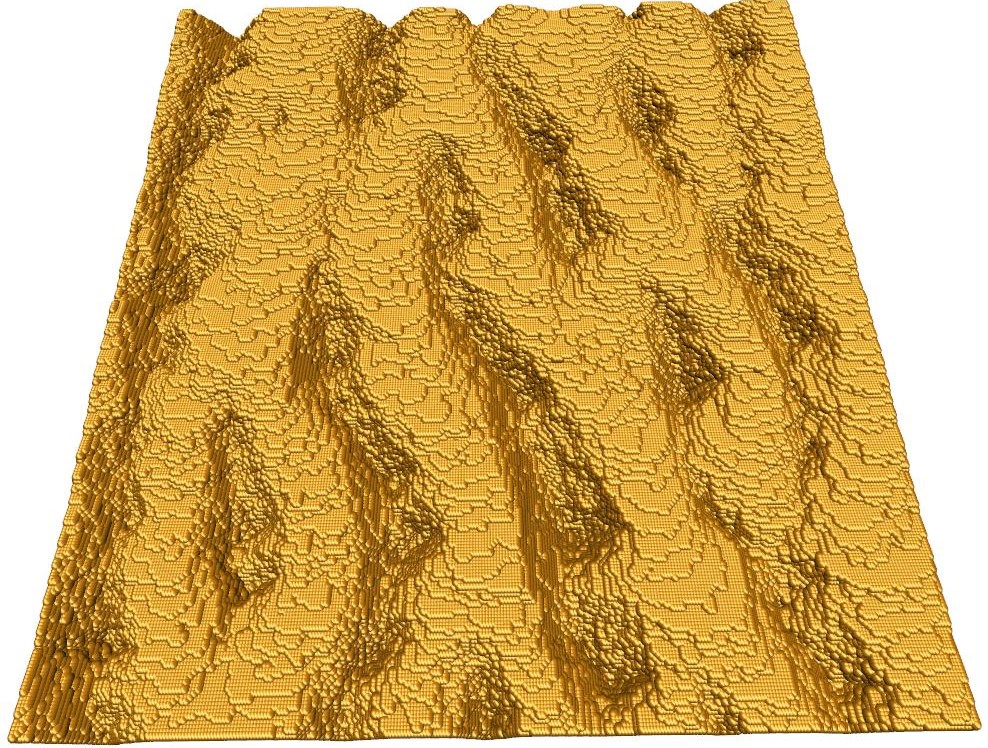}
b)\includegraphics[width=0.3\textwidth]{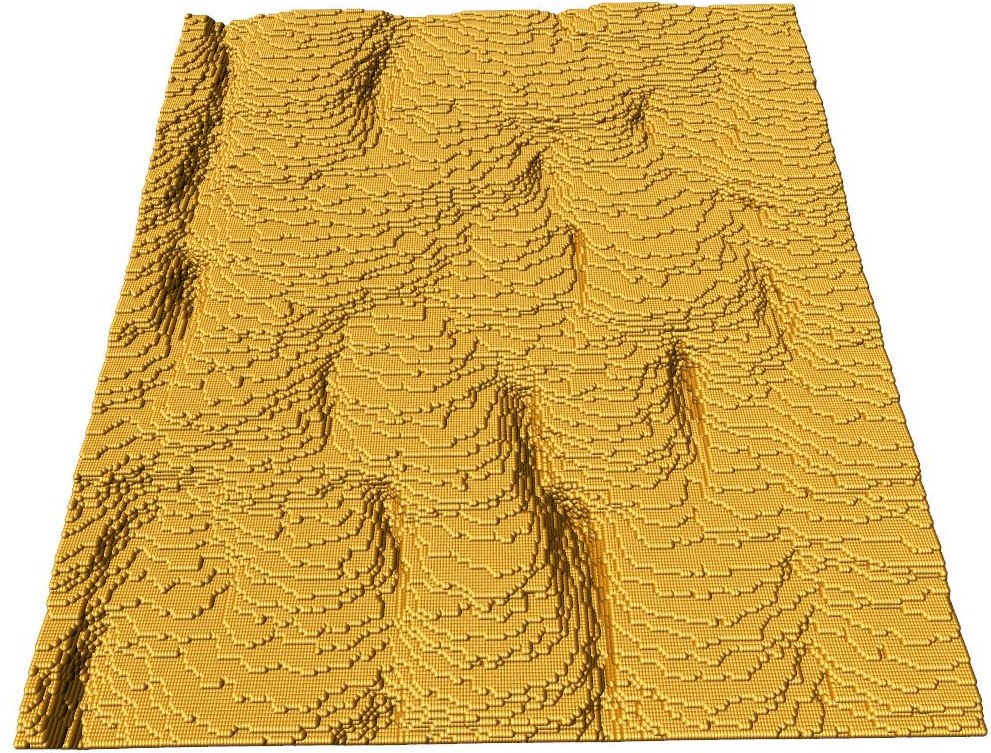}
c)\includegraphics[width=0.3\textwidth]{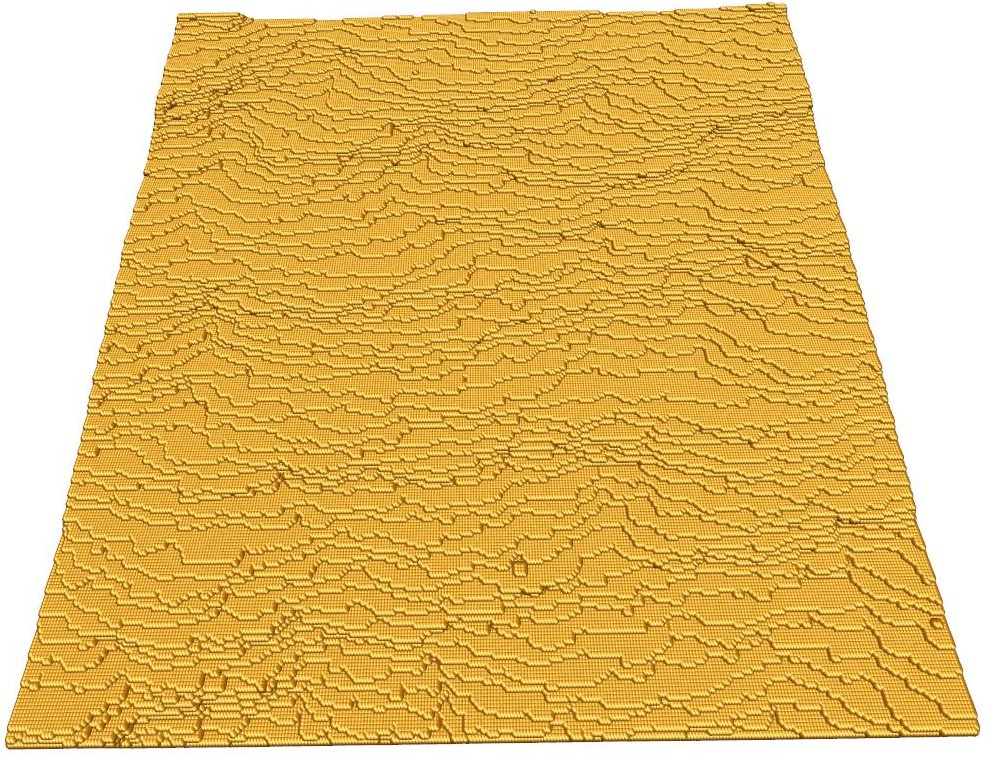}
\caption{Demeandering process: a) The system after $10^6$ growth steps with a flux containing two types of atoms. Along with the neutral atoms, 20\% of additives were introduced, contributing to the presence of potential well $\beta E_{V2}=5$ and ES barrier $\beta E_{ES}=6$ in the system, b) The system after $10^5$ evolution steps with only neutral atoms delivered to the meandered surface, c) The meandered system after $2\cdot10^6$ evolution steps with neutral atoms.}
\label{fig:meander}
\end{figure}
The question now is whether we can smooth the system by eliminating the source of the bunching or meandering process. To test this, after obtaining the bunched structure, we begin supplying the system with only one type of particle. In Figure~\ref{fig:bunch}, the left panel displays the bunches formed in the same system as shown in Fig.~\ref{fig:bunch_form}b after $10^6$ simulation steps. The middle panel shows the pattern obtained after an additional $10^5$ simulation steps, during which only a single type of atom, without additives, was  delivered. The final plot on the right illustrates the outcome of the system  evolution after $2\cdot 10^6$ growth steps with a single component. It is  evident that the crystal continues to grow, and during this process, the step bunches  quickly disintegrate, returning to an even distribution once the cause of their formation is removed.

Let’s explore whether the procedure has a similar effect on meandered structures. We applied the method to the system depicted in Fig.~\ref{fig:meander_form}b, using a slightly smaller system of size $200 \times 200$ while keeping all other parameters identical to those in the previous example. Initially, we generated the meandered structure trough $10^6$ simulation steps (see Fig.~\ref{fig:meander}a). Next, we removed the additives from the incoming flux and allowed a single-component crystal to grow on the meandered structure for $10^5$ additional steps. The resulting configuration is shown in Fig.~\ref{fig:meander}b. The meandered structure appears to smooth out even more rapidly than  the bunched structure. The final plot presents the  system after $2\cdot10^6$ evolution steps with single-component flux, where the surface remained notably smooth.

We also examined whether it’s possible to smooth a surface by adding additives instead of removing them. We tested this approach on both bunched and meandered structures by introducing additives with very weak energy potentials into systems with well-formed surface patterns. In both cases, adding these weak potential additives did indeed result in a smoothing effect on the surface shape. However, this method was effective only when the original structures were not too extensive. It appears that large, three-dimensional formations on the surface are significantly more challenging to smooth out.

\subsection{3D structures}
The examples provided demonstrate that typical surface patterning induced by modifications in surface energy potential due to additives is reversible. Specifically, once a surface patterning mechanism, such as iES barrier for bunches or ES barrier for meanders, is established, it is relatively straightforward to return to a smooth surface by simply removing the source of instability. However, it remains to be seen whether this observation holds true for all types of surface structures formed during crystal growth. Of particular interest is the behavior of nanowires and similar structures. NWs, for instance, are commonly formed on crystal surfaces under liquid metal droplets, with gold often being used for this purpose. Investigating whether these nanowire structures can be similarly smoothed or removed could provide valuable insights into the broader applicability  of this reversal process for various surface structures.

The role that the droplet plays in the development of NWs is to accumulate particles under it, so that at higher concentration they can group and build a crystal structure. This accumulation is confined to the area directly under the droplet. Translating this into the context of surface potentials, the droplet effectively creates a potential well in the region it covers, similar to the potential well illustrated in Fig.~\ref{fig:init_additives}c. This potential well acts as a trap for diffusing atoms, leading to their accumulation and subsequent domain formation. As the NW grows beneath the liquid droplet, the droplet remains positioned atop the NW, causing the process to repeat and the NW to increase in height. This top potential can be accurately modeled using the VicCA approach, as demonstrated in \cite{crystals_MZK,Chabowska-ACS}.

\begin{figure}[hbt]
 \centering
\includegraphics[width=0.42\textwidth]{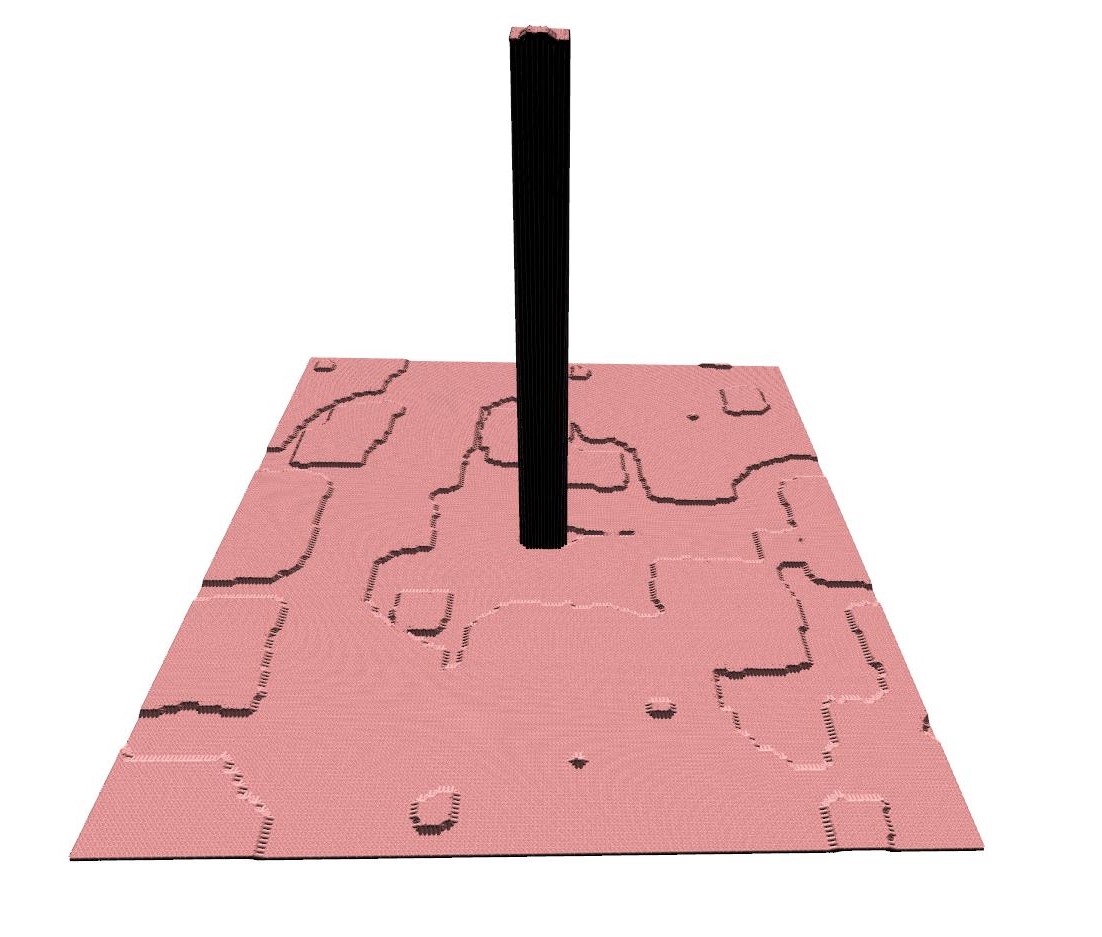}
\includegraphics[width=0.42\textwidth]{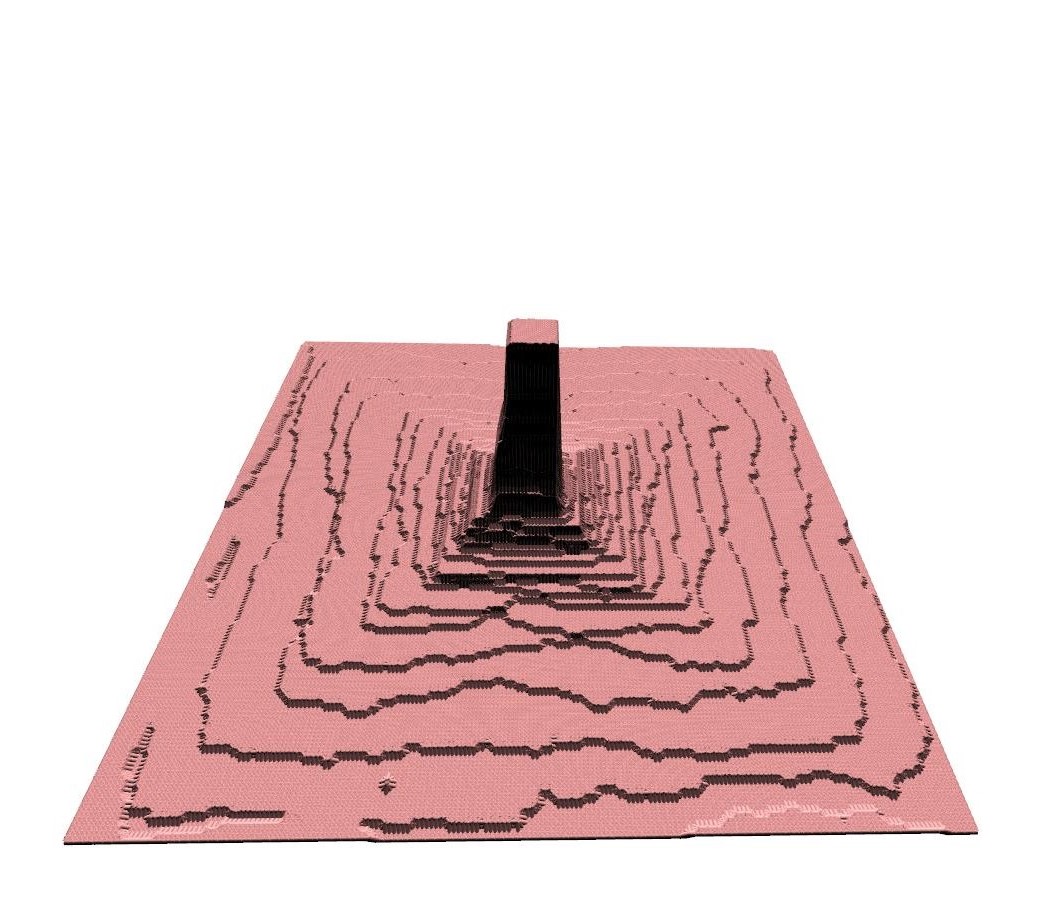}
\caption{The system of a $200\times 200$ lattice sites with initial length of terrace 200, $n_{DS}=10$ and $c_0=0.02$. On the left side, the figure shows the system after $2\cdot10^5$ time steps, during which it evolved under the  influence of a potential energy well placed in the center of system on top of the NW $10\times10$ region. On the right side, the figure presents the system after an additional $4\cdot10^5$ time steps, following the removal of the potential well in the center of the system.}
\label{fig:nw0}
\end{figure}

In Figure~\ref{fig:nw0}, we present simulation results where a potential energy well was positioned at the center of the system. A potential of $\beta E_{V2}=6$ was established within a $10 \times 10$ lattice area and maintained for $2\cdot10^5$ simulation steps. This potential effectively simulates the presence of a gold droplet at the tip of a nanowire. Under these conditions, the NW forms robustly beneath the droplet and continues to grow as long as the potential conditions are upheld. Two types of potential wells were randomly distributed at the base of each step: $\beta E_{V1}=2$ for half of the sites and $\beta E_{V2}=5$ for the remaining sites.

The surface beneath the NW exhibits square-shaped domains. The potential energy well helps concentrate particles at the top, facilitating the nucleation of new layers. A local depression is visible at the foot of the NW, as atoms from the surrounding area migrate to the tip of the NW, where they are adsorbed, fueling its growth. The simulation assumes a high number of diffusion steps, $n_{DS}=10$, corresponding to an effective diffusion coefficient of $10 a^2/t_0$. The growth process was carried out with a relatively high incoming atom flux, defined by $c_0=0.02$. After the initial growth phase, the central potential well was removed, physically representing the removal of the droplet from the apex of the NW. The growth process was then continued using the same parameters. Two notable effects occurred: the NW ceased to grow, and an increasing number of atoms began to attach to its base, forming a square pyramid with the remaining fragment of the NW embedded at its center. The result after $4\cdot10^5$ growth steps is shown in Figure~\ref{fig:nw0} on the right. During this phase of time evolution, the influence of $\beta E_{V2} = 5$ is evident in the rounding of islands. It is clear that further growth, in the absence of the well potential at the NW's tip, begins to smooth the surface; the NW becomes shorter, and a hill forms at its base. This process gradually levels the surface, although complete smoothing would require a significantly longer time. Additionally, various other factors could affect the progression of this smoothing process. It is important to note that in real systems, particularly at considerable NW heights, complete surface smoothing is unlikely to be achieved.

\begin{figure}[hbt]
 \centering
\includegraphics[width=0.425\textwidth]{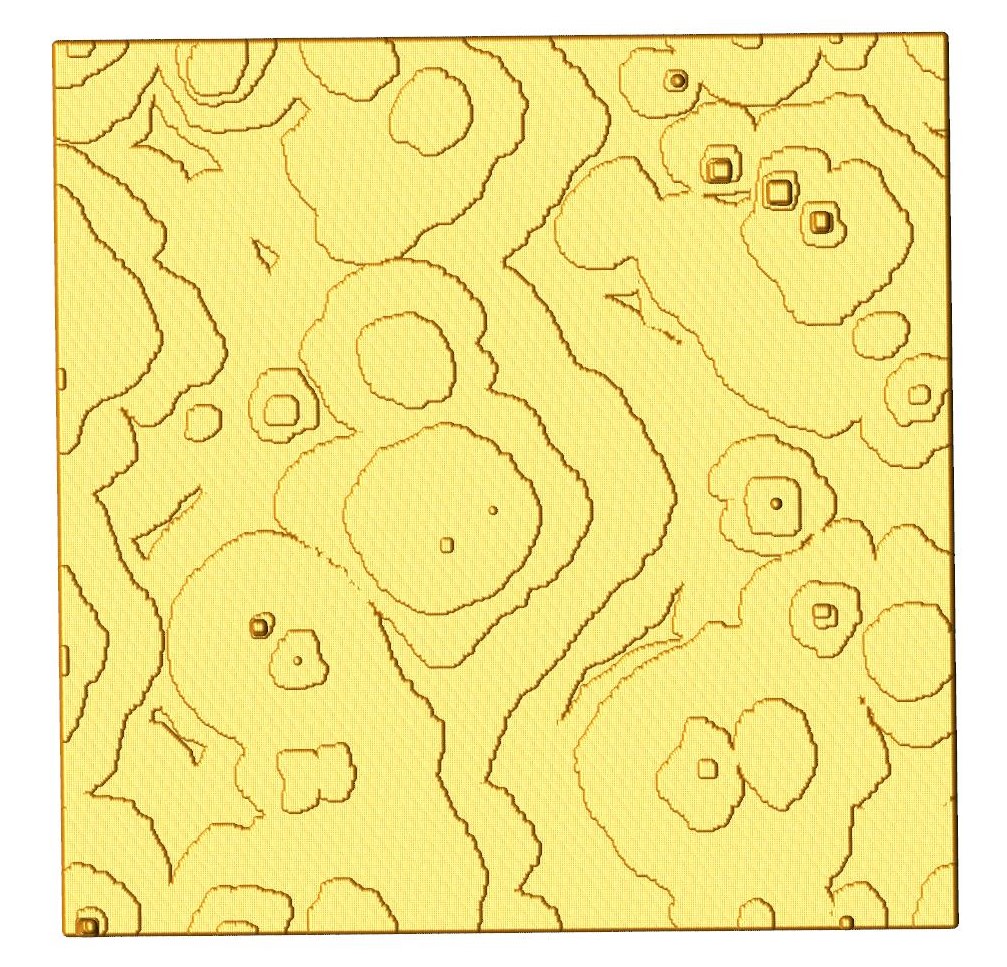}
\includegraphics[width=0.42\textwidth]{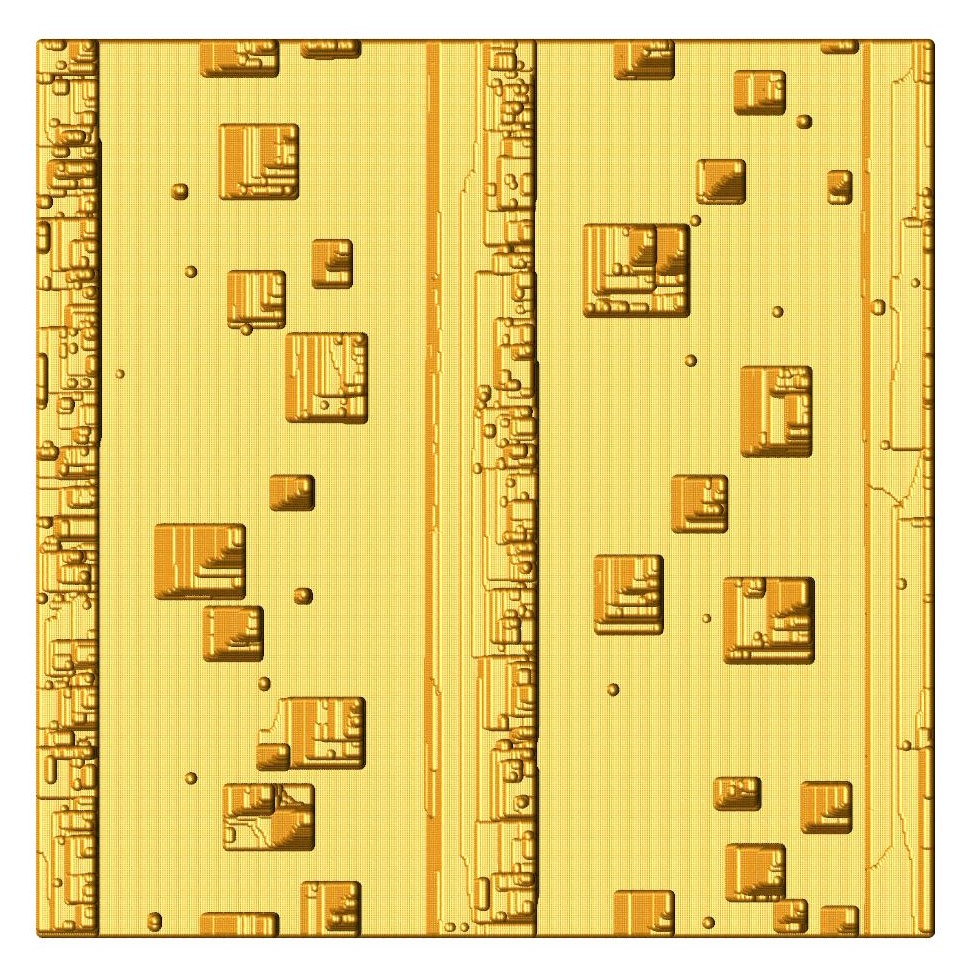}
\includegraphics[width=0.42\textwidth]{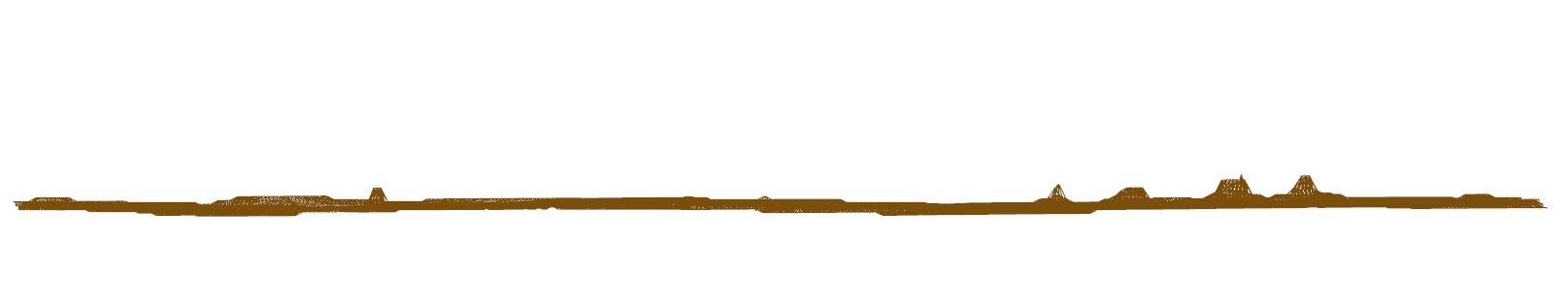}
\includegraphics[width=0.42\textwidth]{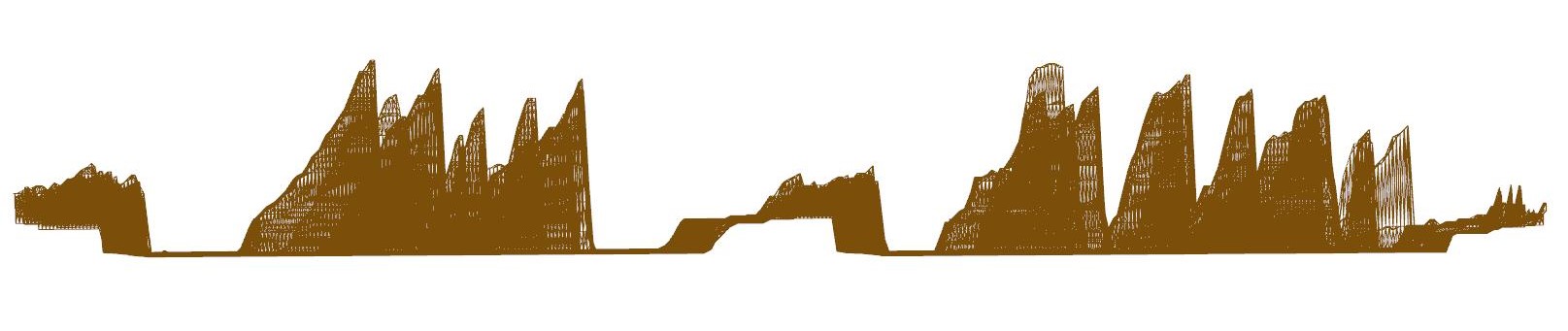}
\caption{The system of $400\times 400$ lattice sites with an initial terrace length of 200, $n_{DS}=6$, $c_0=0.005$. On the left side, both top and side views show the system evolved under a single-component flux of atoms of the first type. On the right side, the system is shown under a flux of atoms of the second type.}
\label{fig:nw11}
\end{figure}
\begin{figure}
 \centering
\includegraphics[width=0.42\textwidth]{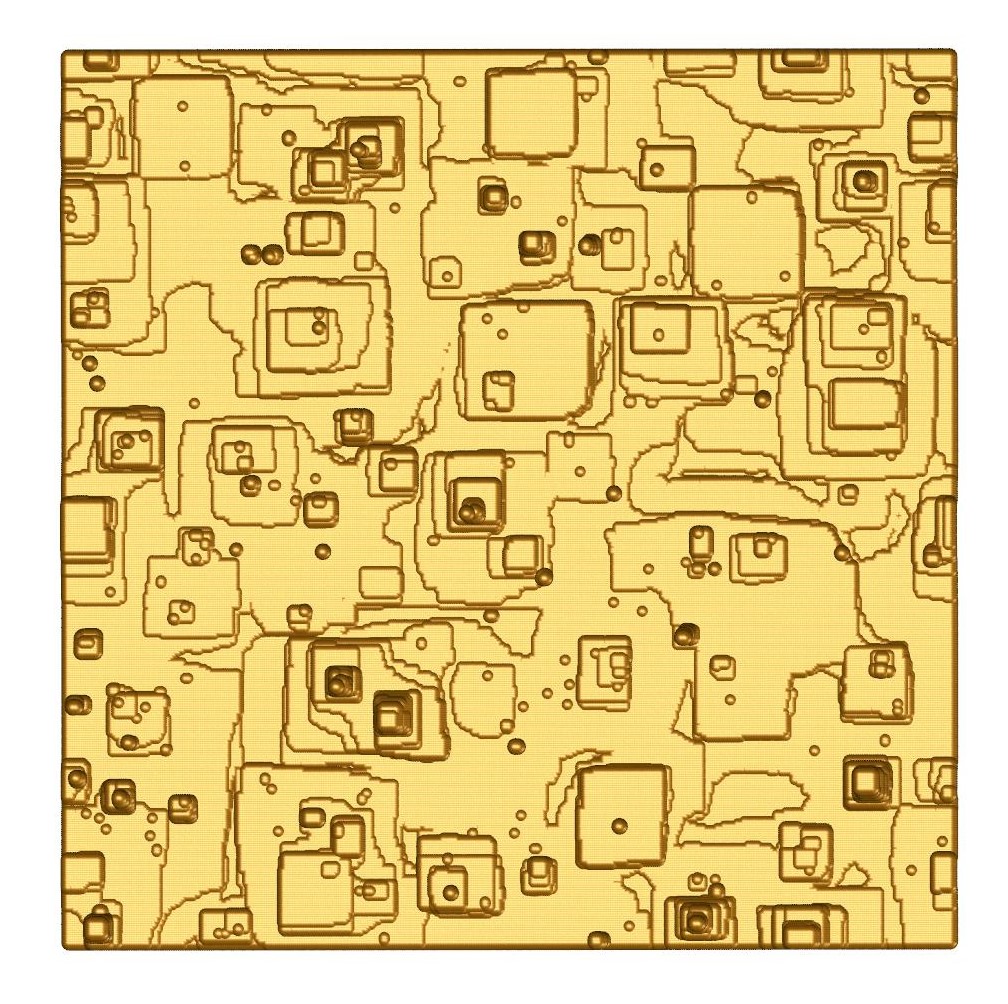}
\includegraphics[width=0.42\textwidth]{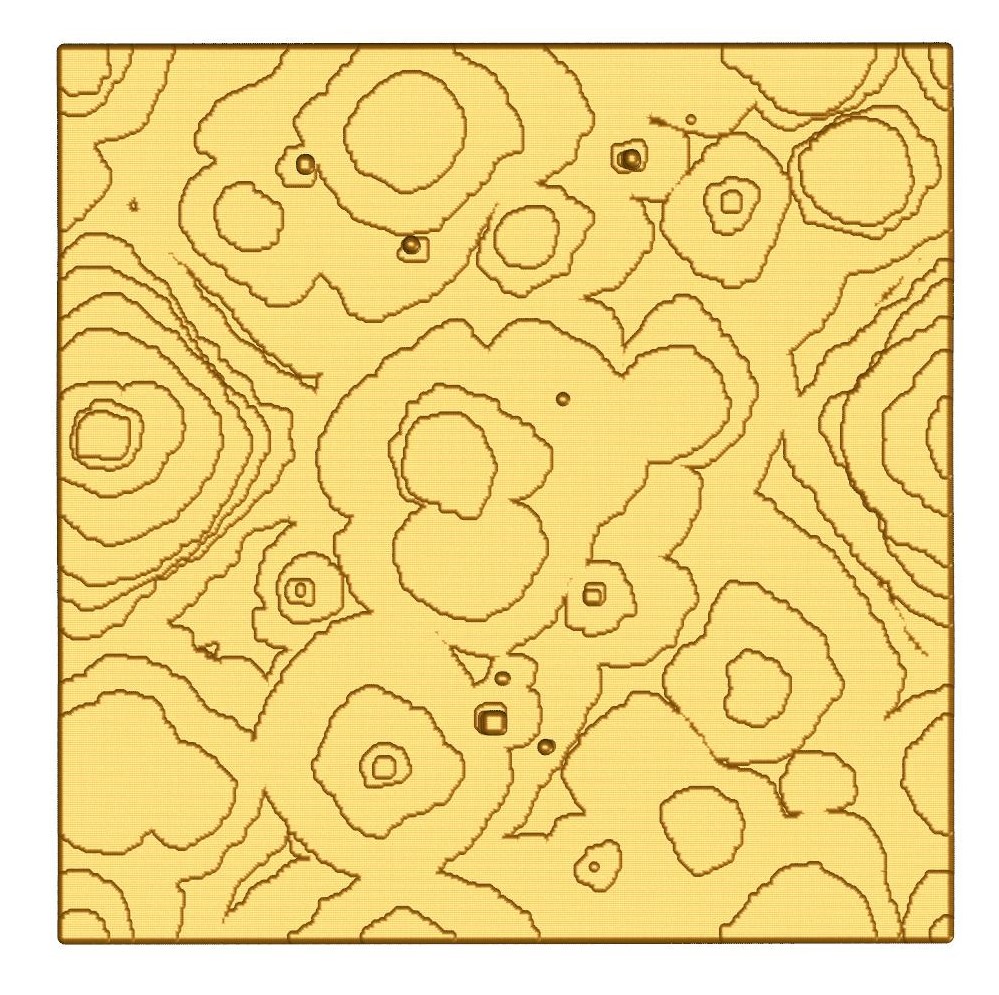}
\includegraphics[width=0.42\textwidth]{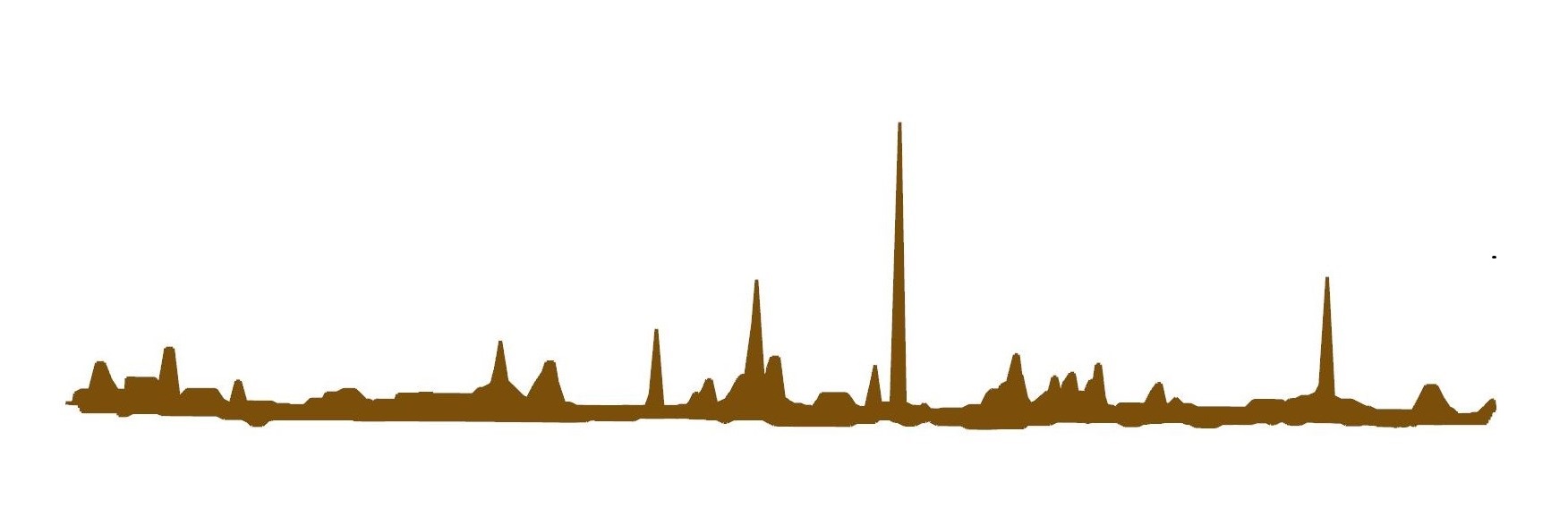}
\includegraphics[width=0.42\textwidth]{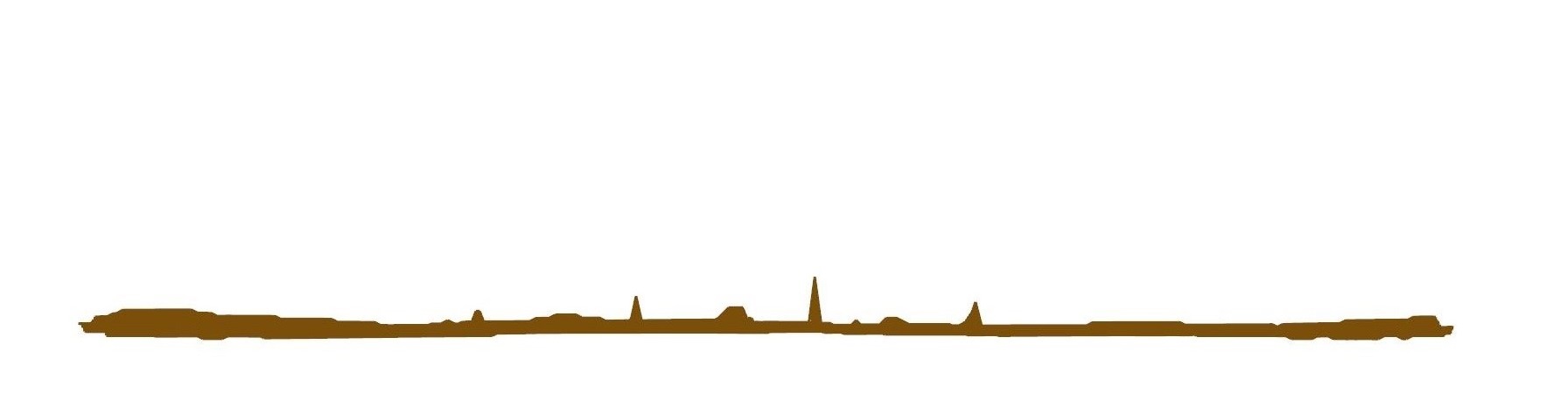}
\caption{The system of $400\times 400$ lattice sites with an initial terrace length of 200, $n_{DS}=6$, $c_0=0.005$ and introduced 60\% of additives. On the left side, both top and side views of the system are shown after $5\cdot10^5$ time steps. On the right side, the system is displayed after an additional $10^6$ time steps under a single-component flux of atoms of the first type.}
\label{fig:nw}
\end{figure}

The example provided demonstrates the formation of NWs beneath a droplet, modeled by a localized potential well. In further studies, we aim to investigate the growth of three-dimensional structures, such as NWs, pyramids, or columns, as a result of atomic movement within a carefully designed potential. We begin with a potential that promotes the formation of meandered structures, specifically a uniform potential with deep energy wells ($\beta E_{V1} = 5$) located below the steps. The system was prepared with wide terraces spanning 200 lattice sites. We reduced the likelihood of nucleation at straight steps by requiring an atom to have two neighbors on both sides for nucleation, instead of just one as before. On the other hand, nucleation on the terraces occurs when a given particle has three neighbors, with a probability of 0.5. This means that, in half of the cases where four particles come together, nucleation takes place on the terrace. A relatively low particle flux is assumed, with parameters $c_0=0.005$ and $n_{DS}=6$. The system is run for $1.5 \cdot 10^6$ simulation time-steps, and the results are presented in Fig.~\ref{fig:nw11}a. It can be observed that this process promotes the formation of large, flat islands. Due to the slow nucleation at steps and the deep potential well, regular, rounded shapes are preferentially formed. Under these conditions, only a few thin and short needles emerge.
We can compare this behavior with the system shown in Fig.~\ref{fig:nw11}b, which exhibits distinctly different characteristics. In this case, a shallow well of depth ($\beta E_{V1} = 2$) is applied, along with an additional ES barrier ($\beta E_{ES} = 2$) at each step. Additionally, it is assumed that each atom adsorbed at a step generates a local potential well of depth ($\beta E_{V2} = 4.95$), which extends to all neighboring sites of the same height. Under these conditions, three-dimensional, NW-like structures can form. As shown in Fig.~\ref{fig:nw11}b, these structures are well-bounded, sharply peaked, and grow significantly above the surface.

We now aim to investigate the behavior of a mixed system, with 40\% of atoms being of the first type and 60\% of the second type. All parameters from the previous examples were retained. Consequently, the potential structure was constructed as follows: the lowest potential is located at the bottom of each step, corresponding to an atom of the first type in the step. A slightly higher, but still low, potential is present at the top of the step, extending to the next two positions and associated with an atom of the second type incorporated at the step. The smallest potential well corresponds to the site next to the position in the step occupied by an atom of the second type.

The system was a $400 \times 400$ lattice, starting with very wide terraces of 200 lattice sites each. The simulation results are depicted in Figure~\ref{fig:nw}, showing the top and side views at two different stages of evolution. On the left, we see the structure after $2\cdot10^5$ simulation steps. On the vicinal surface, two distinct types of structures are visible: flat, square, large islands, and very thin, tall, needle-like nanowires. The simulation was paused at this stage, with the NWs having grown to only about 50 lattice constants in height.
Next, we removed the atoms of second type from the external stream and continued the simulation for another $4\cdot10^5$ steps. The right side of the figure shows that the structures have smoothed out over time, with only remnants of the NWs persisting after this extended growth period. Interestingly, the domains transitioned from square to round shapes, a change attributed to the shift from the potential associated with the atoms of second type to that of the first type.
The potential well at the bottom of the steps is significantly deeper for atoms of the first type compared to those of the second type. As a result, the density of adatoms at the steps increases considerably when the crystal is composed entirely of the first atom type. This reduces the distinction between kink and step positions, leading to the formation of rounded islands. Consequently, the structures grow larger, flatter, and more circular in shape.
This simulation highlights how surface potentials, particularly at step edges, critically influence the crystal growth process and the resulting surface patterns. The transition from square to round islands upon changing the external conditions underscores the complex interplay between atomic interactions and external factors such as additives.

\section{Conclusions}
We conducted an analysis of a process in which semiconductor crystals are modified by substituting some of the atoms in the crystal lattice with additives. This study sheds light on the complex relationship between additives and surface patterns in semiconductor crystals, providing valuable insights into techniques for manipulating and controlling these patterns. To explore this interaction, we used the VicCA model to investigate how the introduction of additives influences the formation of new surface structures. These patterns arise due to the altered interactions and dynamics within the crystal lattice induced by the additives.
In the model, we can precisely modify the effects of additives on the crystal structure by altering the potential energy landscape. By adjusting this potential, we can induce various surface phenomena, such as meandering, bunching, or the formation of more intricate structures like nanowires. Once these new surface structures are formed, they can be smoothed by applying successive layers of a homogeneous composition over the modified crystal. Over time, this process helps to reduce the irregularities introduced by the additives.

Additives can also be employed to smooth pre-existing surface patterns in a homogeneous crystal. However, the effectiveness of this approach depends on the nature of the surface pattern. For simpler patterns, this method can be quite effective, but for more complex, spatially extended patterns like nanowires, applying homogeneous layers is less successful. In such cases, the surface may not be fully smoothed; instead, it may adopt a different shape after treatment. Our study delves into the mechanisms underlying this behavior, exploring how both the presence of additives and the application of homogeneous layers influence the final surface structure of the crystal. This research enhances our understanding of how to control and optimize surface morphologies in semiconductor materials.

\section*{Acknowledgments}
The authors express their gratitude to The Polish National Center for Research and Development (grant no. EIG CONCERT-JAPAN/9/56/AtLv-AlGaN/2023) for providing financial support.
They also extend their thanks to Vesselin Tonchev from the Faculty of Physics at Sofia University, Hristina Popova from Institute of Physical Chemistry of Bulgarian Academy of Sciences and Y. Kangawa from the Research Institute for Applied Mechanics at Kyushu University for their valuable discussions.


\begin{thebibliography}{999}

\bibitem{oreg} Y. Oreg, G. Refael, F. von Oppen, Helical Liquids and Majorana Bound States in Quantum Wires, Phys. Rev. Lett. 105 (2010), 177002. https://doi.org/10.1103/PhysRevLett.105.177002.
\bibitem{lutchyn} R. M. Lutchyn, J. D. Sau, S. Das Sarma, Majorana Fermions and a Topological Phase Transition in Semiconductor-Superconductor Heterostructures, Phys. Rev. Lett. 105 (2010), 077001. https://doi.org/10.1103/PhysRevLett.105.077001.
\bibitem{mourik} V. Mourik, K. Zuo, S. M. Frolov, S. R. Plissard, E. P. A. M. Bakkers, L. P. Kouwenhoven, Signatures of Majorana Fermions in Hybrid Superconductor-Semiconductor Nanowire Devices, Science 336 (2012), 1003–1007.  https://doi.org/10.1126/science.1222360.
\bibitem{grunberg} P. Gr\"{u}nberg, R. Schreiber, Y. Pang, M. B. Brodsky, H. Sowers, Layered Magnetic Structures: Evidence for Antiferromagnetic Coupling of Fe Layers across Cr Interlayers, Phys. Rev. Lett. 57 (1986), 2442. https://doi.org/10.1103/PhysRevLett.57.2442.
\bibitem{baibich} M. N. Baibich, J. M. Broto, A. Fert, V. D. F. Nguyen, F. Petroff, P. Etienne, G. Creuzet, A. Friederich, J. Chazelas, Giant Magnetoresistance of (001)Fe/(001)Cr Magnetic Superlattices, Phys. Rev. Lett. 61 (1988), 2472. https://doi.org/10.1103/PhysRevLett.61.2472.
\bibitem{fermon} C. Fermon, M. Pannetier-Lecoeur, Noise in GMR and TMR Sensors, in: C. Reig, S. Cardoso, S.C. Mukhopadhyay (Eds.), Giant Magnetoresistance (GMR) Sensors. Smart Sensors, Measurement and Instrumentation; Springer: Berlin/Heidelberg, Germany, 2013; Volume 6, pp. 47–70.
\bibitem{zeludev} N. Zheludev, The life and times of the LED — A 100-year history, Nat. Photonics 1 (2007), 189–192. https://doi.org/10.1038/nphoton.2007.34.
\bibitem{yang1} J. J. Yang, M. D. Pickett, X. Li, D. A. A. Ohlberg, D. R. Stewart, R. S. Williams, Memristive switching mechanism for metal/oxide/metal nanodevices, Nat. Nanotechnol. 3 (2008), 429–433. https://doi.org/10.1038/nnano.2008.160.
\bibitem{yang2} J. J. Yang, D. B. Strukov, D. R. Stewart, Memristive Devices for Computing, Nat. Nanotechnol. 8 (2013), 13–24. https://doi.org/10.1038/nnano.2012.240.
\bibitem{verre2012} R. Verre, K. Fleisher, J. F. McGlip, D. Fox, G. Behan, H. Zhang, I. V. Shvets, Controlled in situ growth of tunable plasmonic self-assembled nonoparticle array, Nanotechnol. 23 (2012), 035606. https://dx.doi.org/10.1088/0957-4484/23/3/035606.
\bibitem{yao2016} G. Yao, M. Gao, Y. Ji, W. Liang, L. Gao, S. Zheng, Y. Wang, B. Pang, Y. B. Chen, H. Zeng, H. Li, Z. Wang, J. Liu, C. Chen, Y. Lin, Surface step terrace tuned microstructures and dielectric properties of highly epitaxial CaCu$_3$Ti$_4$O$_{12}$ thin films on vicinal LaAlO$_3$ substrates, Sci. Rep. 6, (2016), 34683. https://doi.org/10.1038/srep34683.
\bibitem{usov2011} V. Usov, C. O. Coileain, I. V. Shvets, Experimental quantitative study into the effects of electromigration field moderation on step bunching instability development on Si(111), Phys. Rev. B 83, (2011), 155321. https://doi.org/10.1103/PhysRevB.83.155321.
\bibitem{shen2007} X. Q. Shen, H. Okamura, Surface step morphologies of GaN films grown on vicinal sapphire (0001) substrate by rf-MBE, J. Cryst. Growth 300, (2007), 75-78. https://doi.org/10.1016/j.jcrysgro.2006.10.206.
\bibitem{omi2005} H. Omi, Y. Homma, V. Tonchev, A. Pimpinelli, New Types of Unstable Step-Flow Growth on Si(111)-(7x7) during Molecular Beam Epitaxy: Scaling and Univeraslity, Phys. Rev. Lett. 95, (2005), 216101. https://doi.org/10.1103/PhysRevLett.95.216101.
\bibitem{ass_turski} H. Turski, F. Krzy\.{z}ewski, A. Feduniewicz-\.{Z}muda, P. Wolny, M. Siekacz, G. Muziol, C. Cheze, K. Nowakowski-Szukudlarek, H. G. Xing, D. Jena, M. Za\l uska-Kotur, C. Skierbiszewski, Unusual step meandering due to Ehrlich-Schwoebel barrier in GaN epitaxy on the N-polar surface, Appl. Surf. Sci. 484, (2019), 771–780. https://doi.org/10.1016/j.apsusc.2019.04.082.
\bibitem{CGD_sawicka} M. Sawicka, H. Turski, K. Sobczak, A. Feduniewicz-Żmuda, N. Fiuczek, O. Gołyga, M. Siekacz, G. Muziol, G. Nowak, J. Smalc-Koziorowska, C. Skierbiszewski, Nanostars in Highly Si-Doped GaN, Cryst. Growth Des. 23, (2023), 5093–5101. https://pubs.acs.org/doi/10.1021/acs.cgd.3c00317
\bibitem{liu1998} D. J. Liu, J. D. Weeks, Quantitative theory of current-induced step bunching on Si(111), Phys. Rev. B 57, (1998), 14891-14900. https://doi.org/10.1103/PhysRevB.57.14891.
\bibitem{pimpinelli2002} A. Pimpinelli, V. Tonchev, A. Videcoq, M. Vladimirova, Scaling and Universality of Self-Organized Patterns on Unstable Vicinal Surfaces, Phys. Rev. Lett. 88, (2002), 206103. https://doi.org/10.1103/PhysRevLett.88.206103.
\bibitem{krug2005} J. Krug, V. Tonchev, S. Stoyanov, A. Pimpinelli, Scaling properties of step bunches induced by sublimation and related mechanisms, Phys. Rev. B 71, (2005), 045412. https://doi.org/10.1103/PhysRevB.71.045412.
\bibitem{tonchev2010} V. Tonchev, B. Ranguelov, H. Omi, A. Pimpinelli, Scaling and universality in models of step bunching: the "C$^+$-C$^-$" model, Eur. Phys. J. B 73, (2010), 539-546. https://doi.org/10.1140/epjb/e2010-00036-3.
\bibitem{stoyanov1997} S. Stoyanov, Current induced step bunching at vicinal surfaces during crystal evolution, Surface Science 370, (1997), 345-354. https://doi.org/10.1016/S0039-6028(96)00966-1.
\bibitem{fk2017_2} F. Krzy\.{z}ewski, M. A. Za\l uska Kotur, Stability diagrams for surface patterns of GaN($000\overline{1}$) as a function of Schwoebel barrier height, J. Crys. Growth 457, (2017), 80-84. https://doi.org/10.1016/j.jcrysgro.2016.04.043.
\bibitem{RMP_Misbah} C. Misbah, O. Pierre-Louis, Y. Saito, Crystal surfaces in and out of equilibrium: A modern view, Rev. Mod. Phys. 82, (2010), 981-1040. https://doi.org/10.1103/RevModPhys.82.981.
\bibitem{Krukowski22} S. Krukowski, K. Sakowski, P. Strak, P. Kempisty, J. Piechota, I. Grzegory, Macrosteps dynamics and the growth of crystals and epitaxial layers, Progress in Crystal Growth and Characterization of Materials 68, (2022), 100581. https://doi.org/10.1016/j.pcrysgrow.2022.100581.
\bibitem{schwoebel1969} R. L. Schwoebel, Step motion on crystal surfaces. II, J. Appl. Phys. 40, (1969), 614-618. https://doi.org/10.1063/1.1657442.
\bibitem{sato2001} M. Sato, M. Uwaha, Growth law of step bunches induced by the Ehrlich-Schwoebel effect, Surface Sci. 493, (2001), 494-498. https://doi.org/10.1016/S0039-6028(01)01258-4.
\bibitem{xie2002} M. H. Xie, S. Y. Leung, S. Y. Tong, What causes step bunching - negative Ehrlich-Schwoebel barrier versus positive incorporation barrier, Surface Sci. 515, (2002), L459-L463. https://doi.org/10.1016/S0039-6028(02)01976-3
\bibitem{schwoebel1966} R. L. Schwoebel, E. J. Shipsey, Step motion on crystal surfaces,  J. Appl. Phys. 37, (1966), 3682-3686. https://doi.org/10.1063/1.1707904.
\bibitem{JAP_krzyzew} F. Krzy\.{z}ewski, M. A. Za\l uska-Kotur, Coexistence of bunching and meandering instability in simulated growth of 4H-SiC(0001) surface, J. Appl. Phys. 115, (2014), 213517. https://doi.org/10.1063/1.4881816.
\bibitem{JK_NL} J.-H. Kang, F. Krizek, M. Za\l uska-Kotur, P. Krogstrup, P. Kacman, H. Beidenkopf, H. Shtrikman, Au-Assisted Substrate-Faceting for Inclined Nanowire Growth, Nano Letters 18, (2018), 4115-4122. https://doi.org/10.1021/acs.nanolett.8b00853.
\bibitem{Arora} S. K. Arora, B. J. O’Dowd, B. Ballesteros, P. Gambardella, I. V. Shvets, Magnetic properties of planar nanowire arrays of Co fabricated on oxidized step-bunched silicon templates, Nanotechnology 23, (2012), 235702. https://doi.org/10.1088/0957-4484/23/23/235702.
\bibitem{AIP_Krasteva} A. Krasteva, H. Popova, F. Krzy\.{z}ewski, M. Za\l uska-Kotur, V. Tonchev, Unstable vicinal crystal growth from cellular automata, AIP Conf. Proc. 1722, (2016), 220014. https://doi.org/10.1063/1.4944246.
\bibitem{Sudoh_2003} K. Sudoh, H. Iwasaki, Step dynamics in faceting on vicinal Si(113) surfaces, J. Phys.: Condens. Matter 15, (2003), S3241. https://doi.org/10.1088/0953-8984/15/47/004.
\bibitem{Neel_2003} N. Néel, T. Maroutian, L. Douillard, H.-J. Ernst, Spontaneous structural pattern formation at the nanometre scale in kinetically restricted homoepitaxy on vicinal surfaces, J. Phys.: Condens. Matter 15, (2003), S3227. https://doi.org/10.1088/0953-8984/15/47/003.
\bibitem{Rahman_2003} T. S. Rahman, A.  Kara, S. Durukanoglu, Structural relaxations, vibrational dynamics and thermodynamics of vicinal surfaces, J. Phys.: Condens. Matter 15, (2003), S3197. https://doi.org/10.1088/0953-8984/15/47/002.
\bibitem{Minoda_2003} H. Minoda, Direct current heating effects on Si(111) vicinal surfaces, J. Phys.: Condens. Matter 15, (2003), S3255. https://doi.org/10.1088/0953-8984/15/47/005.
\bibitem{Rousset_2003} S. Rousset, V. Repain, G. Baudot, Y. Garreau, J. Lecoeur,  Self-ordering of Au(111) vicinal surfaces and application to nanostructure organized growth, J. Phys.: Condens. Matter 15, (2003), S3363. https://doi.org/10.1088/0953-8984/15/47/009.
\bibitem{JCG_krzyzew} F. Krzy\.{z}ewski, M. A. Za\l uska-Kotur, A. Krasteva, H. Popova, V. Tonchev, Step bunching and macrostep formation in 1D atomistic scale model of unstable vicinal crystal growth, J. Cryst. Growth 474, (2017), 135-139. https://doi.org/10.1016/j.jcrysgro.2016.11.121.
\bibitem{prb_Toktarbaiuly} O. Toktarbaiuly, V. O. Usov, C. Coile\'{a}in, K. Siewierska, S. Krasnikov, E. Norton, S. I. Bozhko, V. N. Semenov, A. N. Chaika, B. E. Murphy, O. L\"{u}bben, F. Krzy\.{z}ewski, M. A. Za\l uska-Kotur, A. Krasteva, H. Popova, V. Tonchev, I. V. Shvets, Step bunching with both directions of the current: Vicinal W(110) surfaces versus atomistic-scale model, Phys. Rev. B 97, (2018), 035436. https://doi.org/10.1103/PhysRevB.97.035436.
\bibitem{CGD_krzyzew} F. Krzy\.{z}ewski, M. A. Za\l uska-Kotur, A. Krasteva, H. Popova, V. Tonchev, Scaling and Dynamic Stability of Model Vicinal Surfaces, Cryst. Growth Des. 19, (2019), 821-831. https://doi.org/10.1021/acs.cgd.8b01379.
\bibitem{CGD_popova} H. Popova, F. Krzy\.{z}ewski, M. A. Za\l uska-Kotur, V. Tonchev, Quantifying the Effect of Step--Step Exclusion on Dynamically Unstable Vicinal Surfaces: Step Bunching without Macrostep Formation, Cryst. Growth Des. 20, (2020), 7246-7259. https://doi.org/10.1021/acs.cgd.0c00927.
\bibitem{Popova-CGD23} H. Popova, Analyzing the Pattern Formation on Vicinal Surfaces in Diffusion-Limited and Kinetics-Limited Growth Regimes: The Effect of Step–Step Exclusion, Cryst. Growth Des. 23, (2023), 8875–8888. https://doi.org/10.1021/acs.cgd.3c00952.
\bibitem{crystals_MZK} M. Za\l uska-Kotur, H. Popova, V. Tonchev, Step Bunches, Nanowires and Other Vicinal "Creatures" - Ehrlich–Schwoebel Effect by Cellular Automata, Crystals 11, (2021), 1135. https://doi.org/10.3390/cryst11091135.
\bibitem{Chabowska-ACS} M. A. Chabowska, M. Za\l uska-Kotur, Diffusion-Dependent Pattern Formation on Crystal Surfaces, ACS Omega, 8, (2023), 45779-45786. https://doi.org/10.1021/acsomega.3c06377.
\bibitem{Exp1} A. F. Khokhryakov, Y. N. Palyanov, Y. M. Borzdov, A. S. Kozhukhov, D. V. Sheglov, Influence of a silicon impurity on growth of diamond crystals in the Mg-C system, Diam. and Relat. Mater. 87, (2018), 27–34. https://doi.org/10.1016/j.diamond.2018.05.006.
\bibitem{Exp2} T.-S. Chou, J. Rehm, S. Bin Anooz, Ch. Wouters, O. Ernst, A. Akhtar, Z. Galazka, M. Albrecht, A. Fiedler, A. Popp, Impurity-induced step pinning and recovery in MOVPE-grown (100) $\beta-Ga_2O_3$ film, Appl. Phys. Lett. 126, (2025), 022101. https://doi.org/10.1063/5.0242301.
\bibitem{Exp3} T. P. Menasuta, K. A. Grossklaus, J. H. McElearney, T. E. Vandervelded, Bismuth surfactant enhancement of surface morphology and film quality of MBE-grown GaSb (100) thin films over a wide range of growth temperatures, J. Vac. Sci. Technol. A 42, (2024), 032703. https://doi.org/10.1116/6.0003458.
\bibitem{Exp4} A. Chaney, K. Averett, T. J. Asel, S. Mou, Impact of Ga overpressure on the metal modulation epitaxy growth of AlN/AlGaN short period superlattices, J. Appl. Phys. 137, (2025), 025302. https://doi.org/10.1063/5.0214470.
\bibitem{Exp5} M. J. Siegfried, K.-S. Choi, Elucidating the Effect of Additives on the Growth and Stability of $Cu_2O$ Surfaces via Shape Transformation of Pre-Grown Crystals, J. Am. Chem. Soc. 128, (2006), 10356-10357. https://doi.org/10.1021/ja063574y.
\bibitem{ternary} E. E. Mura, A. Gocalinska, G. Juska, S. T. Moroni, A. Pescaglini, E. Pelucchi, Tuning InP self-assembled quantum structures to telecom wavelength: A versatile original InP(As) nanostructure “workshop”, Appl. Phys. Lett. 110, (2017), 113101. https://doi.org/10.1063/1.4978528.
\bibitem{Th1} T. T. Michely, J. Krug, Islands, Mounds, and Atoms: Patterns and Processes in Crystal Growth Far from Equilibrium, Springer Series in Surface Sciences Vol. 42 Springer-Verlag, Berlin, 2004
\bibitem{Th2} D. Kandel, J. D. Weeks, Simultaneous Bunching and Debunching of Surface Steps: Theory and Relation to Experiments, Phys. Rev. Lett. 74, (1995), 3632. https://doi.org/10.1103/PhysRevLett.74.3632.
\bibitem{Th3} M. Sato, Effect of immobile impurities on motion of steps on a vicinal face, Phys. Rev. E 84, (2011), 061604. https://doi.org/10.1103/PhysRevE.84.061604.
\bibitem{Th4} A. Ben-Hamouda, N. Absi, P. E. Hoggan, A. Pimpinelli, Growth instabilities and adsorbed impurities: A case study, Phys. Rev. B 77, (2008), 245430. https://doi.org/10.1103/PhysRevB.77.245430.
\bibitem{Th5} A. B. H. Hamouda, R. Sathiyanarayanan, A. Pimpinelli, T. L. Einstein, Role of codeposited impurities during growth. I. Explaining distinctive experimental morphology on Cu(001), Phys. Rev. B 83, (2011), 035423. https://doi.org/10.1103/PhysRevB.83.035423.
\bibitem{Chabowska-PRB} M. A. Chabowska, H. Popova, M. Za\l uska-Kotur, Step meandering: The balance between the potential well and the Ehrlich –- Schwoebel barrier, arXiv:2411.12487.  	
https://doi.org/10.48550/arXiv.2411.12487.
\bibitem{akiyama1} T. Ohka, T. Akiyama, A. Muizz Pradipto, K. Nakamura, T. Ito, Effect of step edges on adsorption behavior for GaN(0001) surfaces during metalorganic vapor phase epitaxy: An ab initio study, Cryst. Growth Des. 20, (2020), 4358-4365. https://doi.org/10.1021/acs.cgd.0c00117.
\bibitem{akiyama2} T. Akiyama, T. Ohka, K. Nagai, T. Ito, Effect of step edges on the adsorption behavior on vicinal AlN(0001) surface during metal-organic vapor phase epitaxy: An ab initio study, J. Cryst. Growth 571, (2021), 126244. https://doi.org/10.1016/j.jcrysgro.2021.126244.
\bibitem{akiyama3} T. Akiyama, T. Kawamura, Ab initio study for adsorption behavior on AlN(0001) surface with steps and kinks during metal-organic vapor-phase epitaxy, Jpn. J. Appl. Phys. 63, (2024), 02SP71. https://doi.org/10.35848/1347-4065/ad1896.

\end{thebibliography}

\end{document}